\documentclass[11pt]{article}
 \pdfoutput=1
 \usepackage{jheppub}
 \usepackage[T1]{fontenc} 
 \usepackage{hyperref}
 \usepackage{color}
 \usepackage{caption}
 \usepackage[utf8]{inputenc}
 \usepackage{subcaption}
 \makeatletter
 \def\@fpheader{\relax}
 \makeatother

\subheader{\begin{flushright}
UTTG-07-17
\end{flushright}}

\title{On holographic entanglement entropy of Horndeski black holes}

\author[a]{Elena Caceres,}
\author[a]{Ravi Mohan,}
\author[a,b]{Phuc H. Nguyen}
\affiliation[a]{Theory Group, Department of Physics, University of Texas, Austin, TX 78712, USA}
\affiliation[b]{Maryland Center for Fundamental Physics, University of Maryland, College Park, MD 20742, USA}
\emailAdd{elenac@zippy.ph.utexas.edu}
\emailAdd{ravimohan@utexas.edu}
\emailAdd{pnguye12@umd.edu}

\abstract{We study entanglement entropy in a  particular tensor-scalar theory:  Horndeski gravity. Our goal is  two-fold:  investigate the Lewkowycz-Maldacena proposal for entanglement entropy in the presence of a tensor-scalar coupling  and  address a puzzle existing in the literature regarding the thermal entropy of asymptotically AdS Horndeski black holes. Using the squashed cone method, {\it i.e.}  turning on a conical singularity in the bulk, we  derive the functional for entanglement entropy in Horndeski gravity.  We analyze the divergence structure of the bulk equation of motion. Demanding that the leading divergence of the transverse  component of the  equation of motion vanishes  we identify  the surface where to evaluate the entanglement functional. We show that the  surface obtained is precisely the one that  minimizes said functional. By evaluating the entanglement entropy functional on the horizon we obtain the  thermal entropy for Horndeski black holes; this result clarifies  discrepancies  in the literature. As an application of the functional derived we find the minimal surfaces numerically and study the entanglement plateaux. 
}

\begin{document}

\maketitle

\flushbottom

\section{Introduction}
The replica trick is the  most important technique, if not the only one, to compute entanglement entropy in quantum field theory. In black hole physics, it has been used since the 90s to clarify the relation between the Bekenstein-Hawking entropy and entanglement entropy \cite{Callan:1994py, Susskind:1994sm, Kabat:1995eq, Hotta:1996cq}. It was subsequently applied to study entanglement in conformal field theories \cite{Calabrese:2009qy}, and yielded answers which agree with holographic calculations using the Ryu-Takayanagi formula \cite{Ryu:2006bv,Ryu:2006ef}. In the framework of holography \cite{Maldacena:1997re, Witten:1998qj}, the replica trick has proved to be crucial as a tool to prove the Ryu-Takayanagi formula itself \cite{Lewkowycz:2013nqa}, as well as to extend the formula to cases where the bulk theory of gravity is a higher derivative or higher curvature theory \cite{Dong:2013qoa} \cite{Camps:2013zua}.\\
In this paper, we study the extension of the Ryu-Takayanagi formula in a different direction: the bulk gravity theory includes non-minimally coupled matter. Curiously, while the literature on holographic entanglement in higher curvature theories (such as Lovelock gravity) is already extensive \cite{Dong:2013qoa, Camps:2013zua,deBoer:2011wk, Hung:2011xb, Bhattacharyya:2014yga,Caceres:2015bkr, Bhattacharyya:2013jma}, far less attention has been paid to the case of non-minimal coupling. And the efforts have been concentrated in couplings of the type $ \Phi R$.  From the viewpoint of black hole entropy, non-minimal coupling is an interesting and puzzling topic. The finiteness of the Bekenstein-Hawking entropy seems to indicate that the UV-divergence of the entanglement entropy can be absorbed into Newton's constant. For minimal coupling, this is indeed true, but non-minimal coupling has the potential to spoil this nice structure \cite{Solodukhin:1996jt, Larsen:1995ax}.\\
This paper is an attempt to fill in this gap in the literature. We study a theory with a tensor-scalar coupling of the form $R^{\mu\nu}\partial_\mu\phi \partial_\nu \phi$, a particular case of Horndeski gravity. Horndeski gravity, despite being discovered a few decades ago \cite{Horndeski:1974wa}, fell into oblivion and only received much attention quite recently (for a selection of recent work related to Horndeski gravity, see for example \cite{Kuang:2016edj, Papallo:2017qvl, Jiang:2017imk, Feng:2017jub}). For the purpose of holography, asymptotically AdS black hole solutions of Horndeski gravity have been worked out explicitly \cite{Rinaldi:2012vy, Anabalon:2013oea, Feng:2015oea}, and that the parameter space of the couplings is reasonably well understood in terms of causality and stability \cite{Minamitsuji:2015nca, Kobayashi:2014wsa}.\\

In this paper, we use the squashed cone method to derive the entanglement entropy functional \footnote{See also \cite{Mozaffar:2016hmg} for a different approach to deriving holographic entanglement entropy via field redefinition, which could be applicable to Horndeski gravity.}; we find that it receives a  contribution proportional to the gradient-square of the scalar  field. Determining the precise form of the entanglement functional for Horndeski gravity  is one of the results of this paper. We also show that demanding that the leading divergence of the tranverse, $zz$, component of the equations of motion vanishes implies that the entangling surface is a minimal surface. This is our second result. 

The question of whether the entangling surface minimizes the functional found from the squashed cone method is a topic of recent attention \cite{Dong:2017xht}. Until recently, there exist two main arguments that this should be the case: the argument based on the divergences of the equation of motion, and the cosmic-brane argument. Both these arguments are presented in \cite{Dong:2013qoa} and are based on the equation of motion in the bulk. One of the novelties introduced in \cite{Dong:2017xht} is an argument based directly on the action, without going through the equation of motion. More specifically, \cite{Dong:2017xht} considers a double variation of the bulk action (one with respect to the replica index, and a second one which preserves the strength of the conical defect), and derives the stationarity of the entanglement surface from it.

In higher derivative theories requiring that the surface is  minimal is not enough to cancel all the divergences. This is  because in such  theories  generically all the  components of the field equations diverge, not only the transverse one. Furthermore,  there are subleading divergences not present in Einstein gravity. Horndeski gravity is not a higher derivative theory but we do find a similar pattern of divergences. It is possible that just like in higher derivative theories these divergences cancel if we allow for a more general ansatz \cite{Camps:2014voa}.

One of the original motivations of the present work was related to  a puzzle  existing in the literature of Horndeski black holes. In \cite{Feng:2015oea} the authors perform a careful calculation of the thermal entropy of some particular  Horndeski black holes solutions. They use the Wald formula, Euclidean regularization and the Iyer-Wald formalism and the results do not agree with each other. Evaluating our result for the entanglement entropy  on the horizon we are able to shed some light on this issue. 

The paper is organized as follows: in Section \ref{Sec:Review}, we briefly review the squashed cone method to derive entanglement entropy. In Section \ref{Sec:HEE}, we apply this machinery to Horndeski gravity, and analyze the divergences of the bulk equation of motion. In Section \ref{Sec:ThermalS}, we comment on the thermal entropy and the confusion found in the literature regarding this quantity. In Section \ref{Sec:Numerics}, we find the RT surfaces numerically and study the phase transition from connected surface to disconnected surface in the case of spherical black holes. We conclude in Section \ref{Sec:Conclusion}. We relegate the plots of the RT surfaces to Appendices \ref{App:RTSurfaces} and \ref{App:RTSurfaces2}.

\section{Review of the squashed cone method}\label{Sec:Review}
In this section, we review the squashed cone method, an application of the replica trick to derive holographic entanglement entropy. Besides reviewing the background material, this section also serves to fix the notation for the rest of the paper and to lay out the basic equations to be used later when we apply the formalism to Horndeski gravity. First, recall the microscopic definition of entanglement entropy:
\begin{equation}
    S_{EE} = -\mathrm{Tr}{(\rho \log \rho)}
\end{equation}
where $\rho$ is the reduced density matrix of a subregion of a time-slice of the boundary. To obtain the entanglement entropy, the replica trick tells us to first find the $n^{th}$ R\' {e}nyi entropy, defined by:
\begin{equation}
    S_{n} = -\frac{1}{n-1} \log{\mathrm{Tr}{(\rho^{n})}}
\end{equation}
The entanglement entropy is the analytical continuation of $S_{n}$ as $n \rightarrow 1$: $S_{EE} = \lim_{n \rightarrow 1} S_{n}$. From the path integral representation of $\rho$, we can express $S_{n}$ in terms of the partition function $Z_{n}$ on an appropriate n-sheeted Riemann surface $M_{n}$ as:
\begin{equation}\label{SnZn}
    S_{n} = -\frac{1}{n-1} \left( \log{Z_{n}} - n\log{Z_{1}} \right)
\end{equation}
where $Z_{1}$ is the original partition function. Next, we use the basic holographic relation $Z_{CFT} = e^{-S_{bulk}}$ between the field theory partition function $Z_{CFT}$ and the bulk action $S_{bulk}$. If the boundary is taken to be $M_{n}$, then this relation reads:
\begin{equation}
    Z_{n} = e^{-S[B_{n}]}
\end{equation}
for an appropriate bulk geometry $B_{n}$. Substituting into (\ref{SnZn}), we find:
\begin{equation}
    S_{n} = \frac{1}{n-1} (S[B_{n}]-nS[B_{1}])
\end{equation}
Note that the Riemann surface $M_{n}$ has a discrete $Z_{n}$ symmetry. If we assume that this discrete symmetry also exists for the bulk geometry $B_{n}$, then we can consider the quotient $\hat{B}_{n} = B_{n}/Z_{n}$. Since $B_{n}$ has to be regular in the interior, $\hat{B}_{n}$ is regular except at the fixed points of the $Z_{n}$ symmetry, which now forms a codimension-2 surface around which there is a conical deficit. Moreover, we have:
\begin{equation}
    S[B_{n}] = n S[\hat{B}_{n}]
\end{equation}
where we do not include any contribution from the conical deficit in $S[\hat{B}_{n}]$. The R\'{e}nyi entropy can now be written as:
\begin{equation}
    S_{n} = \frac{n}{n-1} (S[\hat{B}_{n}] - S[B_{n}])
\end{equation}
By taking the limit $n \rightarrow 1$, we find the following formula for the entanglement entropy:
\begin{equation}\label{SquashedConeSEE}
    S_{EE} = \partial_{n} S[\hat{B}_{n}] \bigg|_{n=1} = \partial_{\epsilon} S[\hat{B}_{\epsilon}] \bigg|_{\epsilon=0}
\end{equation}
where we introduced the parameter $\epsilon = 1 - n^{-1}$ which characterizes the strength of the conical deficit. Unlike the original replica index $n$ which only makes sense for integer values, the conical deficit $\epsilon$ can be varied continuously. To summarize: the formula above instructs us to solve the gravity equation of motion with some conical deficit for boundary condition, then evaluate the on-shell action and extract the term first order in $\epsilon$. We stress that the on-shell action does not include any contribution from the conical singularity. In other words, we can integrate over the whole spacetime outside a thin ``tube'' enclosing the conical defect, then at the end shrink the diameter of the tube to zero.\\
On the face of it, formula (\ref{SquashedConeSEE}) involves an integral over the whole spacetime (the on-shell action). It turns out, however, that the entropy only receives contribution from the region near the tip of the cone. This can be seen in several ways. One way is to argue that the variation of the action with $\epsilon$ involves the integral of the equation of motion over all of spacetime, but this integral clearly vanishes. This leaves us with a boundary term near the tip of the cone, which contributes to the entropy. Alternatively, one can also argue as follows. Let us call the action evaluated outside the ``tube'' described above $S_{out}$ and the action inside the ``tube'' $S_{in}$. Since the total action $S_{in} + S_{out}$ is a variation away from an on-shell configuration (due to turning on $\epsilon$), it must be that:
\begin{equation}
\partial_{\epsilon} S_{out} |_{\epsilon=0} = -\partial_{\epsilon} S_{in} |_{\epsilon=0}
\end{equation}
Thus, we can trade the integral outside the tube for the one inside:
\begin{equation}\label{SEEfromSin}
S_{EE} = -\partial_{\epsilon} S_{in} |_{\epsilon = 0}
\end{equation}
Again, the entropy only receives contribution from the near-tip region.\\
The fact that the bulk action localizes to the near-tip region means it is enough to work in an approximate metric near the tip. Such a coordinate system, analogous to Riemann normal coordinates, has been worked out in \cite{Lewkowycz:2013nqa} and \cite{Dong:2013qoa}. The metric takes the form:
\begin{eqnarray}\label{RNC}
ds^{2} &=& e^{2A} [dz d\bar{z} + e^{2A} T (\bar{z}dz - z d\bar{z})^{2}] + (h_{ij} + 2K_{aij}x^{a} + Q_{abij}x^{a}x^{b}) dy^{i}dy^{j} \nonumber \\
&+& 2i e^{2A} (U_{i} + V_{ai}x^{a}) (\bar{z} dz - z \bar{z}) dy^{i} + \dots
\end{eqnarray}
Here use polar-like coordinates $(\rho,\tau)$ or complex coordinates $z = \rho e^{i\tau}$, $\bar{z} = \rho e^{-i\tau}$ for the directions transversal to the surface, and $y^{i}$ for the directions along the surface. The factor $e^{2A}$ is given by: 
\begin{equation}
e^{2A} = (z\bar{z})^{-\epsilon}
\end{equation}
and encodes the conical defect. Also, $K_{zij}$ and $K_{\bar{z}ij}$ is the extrinsic curvature of the surface. Note that we expand to second order in $z$, $\bar{z}$ in the metric (\ref{RNC}). This is sufficient when the gravity equation of motion is second order in the metric, such as Einstein gravity or Horndeski gravity. For such a theory terms at most quadratic in $z$ or $\bar{z}$ in the metric contribute to the curvature (and the on-shell action) at $\rho = 0$. For higher derivative theories, it is in general necessary to keep higher order terms in $z(\bar{z})$ in the metric (\ref{RNC}). The Riemann tensor of the metric (\ref{RNC}) contains the following terms first order in $\epsilon$:
\begin{equation}\label{Rzzbarzzbar}
R_{z\bar{z}z\bar{z}} \sim -\frac{\pi}{2} \epsilon e^{2A} \delta^{2}{(x,y)}
\end{equation}
\begin{equation}\label{Rzizj}
R_{zizj} \sim -\frac{\epsilon}{z} K_{zij}
\end{equation}
\begin{equation}\label{Rzbarizbarj}
R_{\bar{z}i\bar{z}j} \sim -\frac{\epsilon}{\bar{z}} K_{\bar{z}ij}
\end{equation}
The first one above diverges as a (2-dimensional) delta function, and the other two diverge as $z^{-1}$ \footnote{To obtain the delta function, we used the identity $\partial_{z}(1/\bar{z}) = \partial_{\bar{z}}(1/z) = \pi \delta^{(2)}{(x,y)}$. Here $x$ and $y$ are the real and imaginary parts of $z$, respectively.}. Also, the Ricci tensor and Ricci scalar contain the following terms first order in $\epsilon$:
\begin{equation}\label{Rzzbar}
R_{z\bar{z}} = \pi\epsilon \delta^{2}{(x,y)}
\end{equation}
\begin{equation}\label{Rzz}
R_{zz} \sim -\frac{\epsilon}{z} K_{z}
\end{equation}
\begin{equation}\label{Rzbarzbar}
R_{\bar{z}\bar{z}} \sim -\frac{\epsilon}{\bar{z}} K_{\bar{z}}
\end{equation}
\begin{equation}\label{R}
R \sim 4\pi\epsilon e^{-2A} \delta^{2}{(z,\bar{z})}
\end{equation}
Using the formula (\ref{SEEfromSin}) together with the approximate metric (\ref{RNC}), \cite{Dong:2013qoa} derived a formula for holographic entanglement entropy for an arbitrary theory of gravity:
\begin{equation}\label{DongFormula}
    S_{EE} = 2\pi \int d^{d}y \sqrt{h} \left[ \frac{\partial L}{\partial R_{z\bar{z} z \bar{z}}} + \sum_{\alpha} \left( \frac{\partial^{2}L}{\partial R_{zizj} \partial R_{\bar{z}k\bar{z}l} } \right)_{\alpha} \frac{8K_{zij}K_{\bar{z}kl}}{q_{\alpha}+1} \right]
\end{equation}
Here $h$ denotes the induced metric on the surface (we will use $g$ for the bulk metric), $z$ and $\bar{z}$ are complex coordinates transversal to the surface. The first term above is identical to the Wald entropy, except that it is not evaluated on a black hole horizon (or Killing horizon) here. As for the second term, it is an anomaly-like contribution that only arises in theories of gravity quadratic or higher order in the curvature. For Einstein gravity, only the first term above contributes and gives (one quarter of) the area, in agreement with the Ryu-Takayanagi formula.\\
Note that the squashed cone method as described above gives us a functional for entanglement entropy, which we are supposed to evaluate on the surface that is the fixed point of the $Z_{n}$ symmetry in the bulk. In practice finding this surface from its definition is difficult, but - as mentioned in the introduction - there exist general arguments that the surface is also the one minimizing the functional \footnote{It is highly desirable that the surface minimizes the functional, since this would immediately imply that the surface satisfies quantum-information properties of entanglement entropy such as the concavity \cite{Headrick:2007km}.}. One such argument was given in \cite{Lewkowycz:2013nqa} for Einstein gravity, as follows. We go back to the parent space $B_{n}$ by making the period of the angle $\tau$ in the metric (\ref{RNC}) $2\pi n$ instead of $2\pi$, and extract the leading divergence of the Ricci tensor near $z=\bar{z}=0$. The leading divergences of $R_{zz}$ and $R_{\bar{z}\bar{z}}$ are as previously found in (\ref{Rzz}) and (\ref{Rzbarzbar}), but the delta function divergence of $R_{z\bar{z}}$ no longer exists because the parent space is regular at the fixed point of the $Z_{n}$ symmetry. Thus, the $zz$ component of Einstein equation diverges as $\frac{\epsilon}{z}K_{z}$. But, as argued by \cite{Lewkowycz:2013nqa}, we should not expect any divergence on physical grounds. If we demand that the coefficient of the divergence vanishes, then we find $K_{z} = K_{\bar{z}} = 0$. This is precisely the condition of a minimal surface (i.e. one which minimizes the area functional).\\

\section{Entanglement functional for Horndeski gravity}\label{Sec:HEE}
In this section, we apply the squashed cone method to Horndeski gravity. The Lagrangian of the Horndeski theory is \footnote{A remark about convention here is in order. We have multiplied the Lagrangian as written in the papers \cite{Anabalon:2013oea, Feng:2015oea} by an overall factor of $-\frac{1}{16\pi}$. This overall factor has been chosen so that when we plug the Einstein-Hilbert part of the action into (\ref{DongFormula}), we obtain one-quarter of the area times $G^{-1}$.}:
\begin{equation}
    L = -\frac{\kappa}{16\pi} (R - 2\Lambda) + \frac{1}{32\pi} (\alpha g_{\mu\nu} - \gamma G_{\mu\nu}) \partial^{\mu} \chi \partial^{\nu} \chi
\end{equation}
Here $\kappa$ is the inverse Newton's constant ($G^{-1}$). The equation of motion for gravity is:
\begin{eqnarray}
0 &=& \kappa (G_{\mu\nu} + \Lambda g_{\mu\nu}) - \frac{1}{2}\alpha \bigg(\partial_{\mu}\chi \partial_{\nu}\chi - \frac{1}{2}g_{\mu\nu} (\partial \chi)^{2}\bigg) - \frac{1}{2}\gamma \bigg( \frac{1}{2}R \partial_{\mu} \chi \partial_{\nu} \chi - 2 \partial_{\rho}\chi \partial_{(\mu}\chi R_{\nu)}{}^{\rho} \nonumber \\
&-& \partial_{\rho}\chi \partial_{\sigma}\chi R_{\mu}{}^{\rho}{}_{\nu}{}^{\sigma} - (\nabla_{\mu}\nabla^{\rho} \chi)(\nabla_{\nu} \nabla_{\rho} \chi) + (\nabla_{\mu}\nabla_{\nu} \chi) \Box \chi + \frac{1}{2}G_{\mu\nu} (\partial \chi)^{2} \nonumber \\
&-& g_{\mu\nu} \bigg[ -\frac{1}{2}(\nabla^{\rho}\nabla^{\sigma}\chi)(\nabla_{\rho}\nabla_{\sigma}\chi) + \frac{1}{2}(\Box \chi)^{2} - \partial_{\rho}\chi \partial_{\sigma}\chi R^{\rho\sigma} \bigg] \bigg)
\end{eqnarray}
and the one for the scalar field is:
\begin{equation}
\nabla_{\mu} ((\alpha g^{\mu\nu} - \gamma G^{\mu\nu} )\nabla_{\nu}\chi) = 0
\end{equation}
We will derive the functional for holographic entanglement entropy in two ways: the first way is by using  (\ref{DongFormula}), and the second way is by going through the squashed cone method. We will of course obtain the same answer in the end. The reason for this two-pronged derivation is as follows: the formula (\ref{DongFormula}) was technically derived in the absence of matter fields in \cite{Dong:2013qoa}, but is expected to apply even in the presence of matter fields. By deriving the entanglement functional for Horndeski gravity twice, we verify that (\ref{DongFormula}) indeed applies and yields the same answer as the more careful derivation with the matter field included at the outset of the squashed cone method.

\subsection{Derivation from Wald entropy formula}
Since the Horndeski gravity action does not involve terms quadratic or higher order in the curvature, only the Wald-like term in (\ref{DongFormula}) contributes to the entropy, and the anomaly-like term does not. Let us differentiate the Horndeski Lagrangian with respect to the Riemann tensor:
\begin{eqnarray}\label{dLdR}
    \frac{\partial L}{\partial R_{\mu\nu\rho\sigma}} &=& - \frac{\kappa}{32\pi} (g^{\mu\rho}g^{\nu\sigma} - g^{\nu\rho}g^{\mu\sigma}) - \frac{\gamma}{128\pi} [ g^{\mu\rho}\chi^{,\nu}\chi^{,\sigma} - g^{\nu\rho}\chi^{,\mu}\chi^{,\sigma} + g^{\nu\sigma}\chi^{,\mu}\chi^{,\rho} - g^{\mu\sigma}\chi^{,\nu}\chi^{,\rho} \nonumber \\
    &-& (g^{\mu\rho}g^{\nu\sigma} - g^{\nu\rho}g^{\mu\sigma}) \chi^{,\lambda} \chi_{,\lambda} ]
\end{eqnarray}
where $\chi^{,\alpha} = g^{\alpha\beta} \chi_{,\beta}$. We need to take the $z\bar{z}z\bar{z}$ component of the expression above (in the coordinate system of the metric \ref{RNC}) and evaluate on the surface (where $z=\bar{z}=0$). It is enough to truncate the metric \ref{RNC} to zeroth order in $z (\bar{z})$ (with the conical defect turned off): 
\begin{equation}
ds^{2} = dzd\bar{z} + h_{ij} dy^{i}dy^{j}
\end{equation}
Upon substituting the metric components into (\ref{dLdR}), the partial derivative $\partial L / \partial R_{z\bar{z}z\bar{z}}$ simplifies to:
\begin{equation}
\frac{\partial L}{\partial R_{z\bar{z}z\bar{z}}} = \frac{\kappa}{8\pi} - \frac{\gamma}{32\pi} \left( \chi^{,\lambda} \chi_{,\lambda} - \chi^{,z} \chi^{\bar{,z}} \right)
\end{equation}
Upon further expanding $\chi^{,\lambda}\chi_{,\lambda} = \chi^{,z}\chi^{,\bar{z}} + h_{ij}\chi^{,i}\chi^{,j}$, this further simplifies to:
\begin{equation}
\frac{\partial L}{\partial R_{z\bar{z}z\bar{z}}} = \frac{\kappa}{8\pi} - \frac{\gamma}{32\pi} h_{ij} \chi^{,i} \chi^{,j}
\end{equation}
The functional for holographic entanglement entropy then reads:
\begin{equation}\label{Functional}
S_{EE} = \frac{\kappa}{4} \int d^{d}y \sqrt{h} \left[ 1 - \frac{\gamma}{4 \kappa} h_{ij} \chi^{,i} \chi^{,j} \right]
\end{equation}
Thus, the entropy receives a Wald-like correction proportional to the norm-squared of the gradient of the scalar field on the surface. Equation \ref{Functional} is one of the results of this paper. By analogy with the RT formula, the entanglement functional is expected to be evaluated in the surface that minimizes its value. In subsection \ref{Divergences} we will show that this is indeed the right thing to do. More precisely,  we will show that  demanding that the divergences of the equations of motion  cancel yields  a condition that is exactly the equation of the surface that minimizes \ref{Functional}!. This will be another important result of our  work.  \\

\subsubsection{Minimization}
Let us now explicitly minimize the functional (\ref{Functional}) to derive the equation characterizing the surface (the ``surface equation''). To do this, we vary the embedding functions $x^{\mu}{(y^{i})}$ of the surface (where $x^{\mu}$ denotes the bulk coordinates and $y^{i}$ denotes the coordinates on the surface), and compute the variation $\delta S$ of the functional due to $\delta x^{\mu}$. When the embedding is varied, the value of the functional is varied due to two effects: (1) the change of the induced metric, and (2) the change of the value of the scalar field on the surface. Thus, we have:
\begin{equation}\label{deltaFunctional}
    \delta S = \int d^{d}y \left( \frac{\delta \mathcal{L}}{\delta h_{ij}} \delta h_{ij} + \frac{\delta \mathcal{L}}{\delta \chi} \delta \chi \right)
\end{equation}
with $\delta h_{ij}$ and $\delta \chi$ related to $\delta x^{\mu}$ as:
\begin{equation}
\delta h_{ij} = g_{\mu\nu} \left( \frac{\partial \delta x^{\mu}}{\partial y^{i}} \frac{\partial x^{\nu}}{\partial y^{j}} + \frac{\partial x^{\mu}}{\partial y^{i}} \frac{\partial \delta x^{\nu}}{\partial y^{j}} \right)
\end{equation}
\begin{equation}
\delta \chi = \frac{\partial \chi}{\partial x^{\mu}} \delta x^{\mu}
\end{equation}
We now adopt the coordinate system $x^{\mu} = (y^{i},z,\bar{z})$ of the metric (\ref{RNC}). This coordinate system has the nice feature that it splits into coordinates on the surface ($y^{i}$) and coordinates transversal to the surface ($z$, $\bar{z}$). It is enough to consider the variations $\delta z$ and $\delta \bar{z}$ (i.e. in the normal direction), since these actually change the embedding, whereas the variations $\delta y^{i}$ merely correspond to coordinate transformations on the surface and should not change the value of the functional. Under the variation $\delta z$ ($\delta \bar{z}$), the first term in (\ref{deltaFunctional}) can be cast in terms of the extrinsic curvature of the surface as:
\begin{equation}
\frac{\delta \mathcal{L}}{\delta h_{ij}} \delta h_{ij} = 2 K_{a ij}\frac{\delta L}{\delta h_{ij}} \delta a
\end{equation}
where $a=z,\bar{z}$ (see for example \cite{Bhattacharyya:2013jma} for more details). For our specific functional (\ref{Functional}), this becomes after some algebra:
\begin{equation}
\frac{\delta \mathcal{L}}{\delta h_{ij}} \delta h_{ij} = \frac{\sqrt{h}}{4G} \left[ \left( 1 - \frac{\gamma G}{4} h^{ij} \chi_{,i} \chi_{,j} \right) K_{a} + \frac{\gamma G}{2} K_{a}^{ij} \chi_{,i} \chi_{,j} \right] \delta a
\end{equation}
Next, consider the second term in (\ref{deltaFunctional}). Under the $\delta z$ ($\delta \bar{z}$), we have:
\begin{equation}
\frac{\delta \mathcal{L}}{\delta \chi} \delta \chi = \frac{\gamma}{8} \sqrt{h} \left( h^{ij} D_{i} D_{j} \chi \right) \chi_{,a} \delta a
\end{equation}
where $D_{i}$ is the covariant derivative on the surface. Putting both contributions of $\delta S$ together, and equating the result to zero, we finally find the surface equation:
\begin{equation}\label{SurfaceEqn}
\left( 1 - \frac{\gamma G}{4} h^{ij} \chi_{,i} \chi_{,j} \right) K_{a} + \frac{\gamma G}{2} K_{a}^{ij} \chi_{,i} \chi_{,j} + \frac{\gamma G}{2} \chi_{,a} h^{ij} D_{i}D_{j}\chi = 0
\end{equation}
In particular, for $\gamma=0$ we recover the minimal surface condition $K_{a} = 0$ of Einstein gravity. Note also that the horizon of a static, spherically symmetric black hole solution also satisfies the surface equation above, assuming that it is also a Killing horizon (as is the case with more familiar black holes). Indeed, for a Killing horizon the extrinsic curvature $K_{ij}$ vanishes, and spherical symmetry also implies that the scalar field is constant on the horizon, so that $D_{i}\chi$ vanishes.\\
Thus, the thermal entropy of the black hole should be found by evaluating the functional (\ref{Functional}) on the horizon. An immediate interesting consequence of this is that the thermal entropy is equal to the usual Bekenstein-Hawking entropy (assuming again spherical symmetry):
\begin{equation}\label{SthermalHorndeski}
S_{\mathrm{thermal}} = \frac{A}{4 G}
\end{equation}
where $A$ is the area of the horizon. Indeed, spherical symmetry implies that the Wald-like correction in (\ref{Functional}) simply vanishes.\\
As pointed out in \cite{Feng:2015oea}, however, the thermal entropy of Horndeski black holes is surprisingly subtle, with different computation methods yielding different answers. We will revisit this issue in Section \ref{Sec:ThermalS}, but we assume for the rest of this paper that (\ref{SthermalHorndeski}) is the correct one. We also stress that the Wald-like correction only vanishes for the thermal entropy, and not for entanglement entropy of a boundary subregion as in general the scalar field is not constant on the Ryu-Takayanagi surface.

\subsection{Re-derivation by squashed cone method}\label{SquashedCone}
We proceed to rederive the functional (\ref{Functional}) using the squashed cone method, with the scalar field put in at the outset. We will need to know the expansion of the scalar field near the surface, in the coordinate system of the metric (\ref{RNC}). This expansion is $\epsilon$-dependent, since the scalar field has to readjust itself to the the conical defect, and this $\epsilon$ dependence has to be fed into the on-shell action to see if it gives rise to any additional term compared to the functional (\ref{Functional}). In the end, we will find that there are no additional terms, thus the expectation that the functional (\ref{Functional}) applies even with matter fields is borne out.\\
Let us first recall how the coordinate system of (\ref{RNC}) is constructed. In the parent space $B_{n}$ (i.e. before the quotienting by $Z_{n}$), we set up polar-like coordinate $(\tilde{\rho},\tilde{\tau})$ in the plane transversal to the surface. Since the bulk has to be regular at the surface and also has the replica $Z_{n}$ symmetry, any $\tilde{\tau}$ dependence has to be through $\tilde{\rho}^{n} e^{\pm i n \tilde{\tau}}$. In the quotient $\hat{B}_{n}$, we redefine coordinates as follows:
\begin{equation}\label{QuotientCoordrho}
\rho = \left( \frac{\tilde{\rho}}{n} \right)^{n}
\end{equation}
\begin{equation}\label{QuotientCoordtau}
\tau = n \tilde{\tau}
\end{equation}
The metric in terms of $\rho$ and $\tau$ then takes the form (\ref{RNC}). For the scalar field, we can proceed similarly. Regularity and the replica symmetry dictate that the scalar field near the surface is a function of $\tilde{\rho}^{2}$ and $\tilde{\rho}^{n} e^{\pm i n \tau}$ in the parent space $B_{n}$. In the quotient space, we find - to second order in $z$ or $\bar{z}$ - a rather complicated expansion:
\begin{eqnarray}\label{chiRNC}
    \chi &=& [\chi_{0;0} + \chi_{0;1}(z\bar{z})^{\epsilon} + \chi_{0;2}(z\bar{z})^{2\epsilon} + \dots] 
    + [\chi_{z;0} + \chi_{z;1}(z\bar{z})^{\epsilon} + \chi_{z;2}(z\bar{z})^{2\epsilon} + \dots ]z \nonumber \\
    &+& [\chi_{\bar{z};0} + \chi_{\bar{z};1}(z\bar{z})^{\epsilon} + \chi_{\bar{z};2}(z\bar{z})^{2\epsilon} + \dots ]\bar{z} + [\chi_{z^{2};0} + \chi_{z^{2};1}(z\bar{z})^{\epsilon} + \chi_{z^{2};2}(z\bar{z})^{2\epsilon} + \dots ]z^{2} \nonumber \\
    &+& [\chi_{\bar{z}^{2};0} + \chi_{\bar{z}^{2};1}(z\bar{z})^{\epsilon} + \chi_{\bar{z}^{2};2}(z\bar{z})^{2\epsilon} + \dots ]\bar{z}^{2} + [\chi_{z\bar{z};0} + \chi_{z\bar{z};1}(z\bar{z})^{\epsilon} + \chi_{z\bar{z};2}(z\bar{z})^{2\epsilon} + \dots ](z\bar{z})^{1-\epsilon} \nonumber \\
    &+& \dots
\end{eqnarray}
where the coefficients $\chi_{0;0}$, $\chi_{0;1}$ etc. are functions of $y^{i}$. Let us explain how the different powers in the expansion above arise, especially the subleading terms in each of the square brackets above. Consider for example the term with $\chi_{0;1}$. This term comes from the power $\tilde{\rho}^{2(n-1)}$ in the parent space, which is consistent with regularity and replica symmetry in the bulk. Similarly, the power $(\tilde{\rho}^{2})^{2n-2}$ gives rise to $(z\bar{z})^{2\epsilon}$ etc. Note that at $\epsilon=0$, each square bracket above collapses to a constant and we find:
\begin{equation}\label{chiRNCepsilon0}
    \chi = \chi_{0} + \chi_{z}z + \chi_{\bar{z}}\bar{z} + \chi_{z^{2}}z^{2} + \chi_{\bar{z}^{2}}\bar{z}^{2} + \chi_{z\bar{z}} z\bar{z} \dots
\end{equation}
with $\chi_{0} = \sum_{k=0}^{\infty} \chi_{0;k}$, $\chi_{z} = \sum_{k=0}^{\infty} \chi_{z;k}$ and so on. In other words, each term in the series at $\epsilon=0$ is a whole infinite series at $\epsilon \neq 0$. This is sometimes dubbed the ``splitting problem'' \cite{Dong:2017xht, Miao:2014nxa}.\\
Before proceeding with the squashed cone method, we would like to make two remarks about the expansion (\ref{chiRNC}) for the sake of clarity. First, the coefficients $\chi_{0;0}$ etc. in this expansion are taken to be independent of $\epsilon$. This is not strictly true; for example, the term $\tilde{\rho}^{n}e^{ i n \tilde{\tau}}$ in the parent space becomes $n^{n}z \approx (1 + \epsilon + \mathcal{O}{(\epsilon^{2})})z$ after the coordinate transformation (\ref{QuotientCoordrho}) and (\ref{QuotientCoordtau}). For general $n$, the $\epsilon$ corrections of the coefficients are not ignorable. However, throughout this paper we work with the $\epsilon \approx 0$ (or $n \approx 1$) regime, and these $\epsilon$ corrections are small. Moreover, even if we keep the $\epsilon$ corrections, they will not contribute to the entanglement entropy anyway. Note also that this discarding of the $\epsilon$ corrections is not specific to the scalar field expansion: the same treatment was applied to the metric expansion (\ref{RNC}), i.e. coefficients in this expansion in general receive $\epsilon$ correction but they have been discarded for the reasons above.\\
Our second remark is that the expansion (\ref{chiRNC}) implies that the entanglement entropy functional depends on the scalar field through the coefficients $\chi_{0;0}$, $\chi_{0;1}$, $\chi_{z;0}$, $\chi_{z;1}$. This may appear peculiar for the following reason. Our experience with the usual Ryu-Takayanagi formula for Einstein gravity, as well as the Jacobson-Myers functional for Gauss-Bonnet gravity, would suggest that the entanglement functional should be a function of the value of the scalar field or its derivatives at $\epsilon=0$. On the other hand, each of the coefficient $\chi_{0;0}$, $\chi_{0;1}$ etc. does not really have an intrinsic meaning at $\epsilon=0$ (only the sums $\sum_{k} \chi_{0;k}$, $\sum_{k} \chi_{z,k}$  etc. do: the first is the value of the scalar field on the minimal surface, and the second is the derivative of the scalar field in the $z$ direction on the minimal surface).\\
Fortunately, we find that the entanglement functional indeed depends on the coefficients in (\ref{chiRNC}) only through the sums $\sum_{k} \chi_{0;k} = \chi_{0}$, $\sum_{k} \chi_{z;k} = \chi_{z}$ etc., in other words the functional only depends on quantities which ``make sense'' at $\epsilon=0$. Roughly speaking, this is because we need to look for terms first order in $\epsilon$ in the action. While one can extract many terms first order in $\epsilon$ from the expansion (\ref{chiRNC}), the only terms first order in $\epsilon$ that matter come from the curvature at the tip of the cone, i.e. the coupling of the scalar field to the conical singularity. But the scalar field that appear in such terms can be evaluated at $\epsilon=0$ since the curvature is already first order in $\epsilon$, therefore only the expansion (\ref{chiRNCepsilon0}) really matters.\\
Using the scalar field expansion together with equations (\ref{Rzz}), (\ref{Rzzbar}), (\ref{Rzbarzbar}) and (\ref{R}) for the singular terms of the Ricci tensor and Ricci scalar, we then find the term first order in $\epsilon$ of the non-minimal coupling part of the action:
\begin{eqnarray}\label{SNMC}
S_{NMC}^{(1)} &=& \frac{\gamma\epsilon}{32\pi} \int_{\rho \sim 0} \sqrt{g} d^{d}x \bigg[ 8 e^{-4A} \left( \frac{K_{z}}{z} (\chi_{\bar{z}})^{2} + \frac{K_{\bar{z}}}{\bar{z}} (\chi_{z})^{2} \right) \\ \nonumber
&+& 2\pi e^{-2A} \delta^{2}{(x^{1},x^{2})} h^{ij}\chi_{0,i}\chi_{0,j}  \bigg]
\end{eqnarray}
where the subscript $NMC$ stands for non-minimal coupling, and the superscript $(1)$ means first order in $\epsilon$. We can easily argue that the terms in $S_{NMC}^{(1)}$ containing $K_{\bar{z}}$ and $K_{z}$ vanish. This is because the integral is regular at the origin and we integrate over an infinitesimally small region around the tip of the cone. Thus, we are left with only the term with the delta function. The above then simplifies to:
\begin{equation}
S^{(1)}_{NMC} = \frac{\gamma\epsilon}{16} \int d^{d-2}y \sqrt{h} h^{ij}\chi_{,i}\chi_{,j}
\end{equation}
where we have replaced $\chi_{0,i}$ by $\chi_{,i}$ since the functional above is to be evaluated on the surface $z=\bar{z}=0$, where those two quantities agree. By the formula (\ref{SEEfromSin}), this contributes to the entropy an amount:
\begin{equation}
S_{EE}^{(NMC)} = -\frac{\gamma}{16} \int \sqrt{h} d^{d-2}y h^{ij} \chi_{,i} \chi_{,j}
\end{equation}
We now combine the contribution above to the area functional from the Einstein-Hilbert part of the action, to obtain the total functional:
\begin{equation}\label{HEEHorndeski}
S_{EE} = \frac{1}{4G} \int d^{d-2}y \sqrt{h} \left(1 - \frac{\gamma G}{4} h^{ij} \partial_{i}\chi \partial_{j}\chi \right)
\end{equation}
in agreement with the functional previously derived from \eqref{DongFormula} which in this case is equivalent to   the Wald entropy formula.

\subsection{Canceling divergences of the equation of motion}\label{Divergences}
In this subsection, we go back to the parent space $B_{n}$ and extract the leading order divergence in the equation of motion. We will continue to work with the metric (\ref{RNC}) but with $\tau \sim \tau + 2\pi n$ so that there is no conical singularity. Similarly, we will use the expansion (\ref{chiRNC}) of the scalar field, but this expansion is now understood as an expansion in the parent space rather than the quotient space.\\
The divergences of the bulk equation of motion come from two contributions: (1) first, there are divergences from the components $R_{zizj}$ and $R_{\bar{z}i\bar{z}j}$ of the Riemann tensor,  like in Einstein gravity; (2) Secondly, the components $\nabla_{z}\nabla_{z}\chi$ and $\nabla_{\bar{z}} \nabla_{\bar{z}} \chi$ of the second covariant derivative of the scalar field also contribute divergences. This is because the Christoffel symbols $\Gamma^{z}_{zz}$ and $\Gamma^{\bar{z}}_{\bar{z}\bar{z}}$ also contain $1/z$ divergences:
\begin{equation}
\Gamma^{z}_{zz} \sim -\frac{\epsilon}{z} 
\end{equation}
\begin{equation}
\Gamma^{\bar{z}}_{\bar{z}\bar{z}} \sim -\frac{\epsilon}{\bar{z}}
\end{equation}
 Therefore, the following components of $\nabla_{\mu} \nabla_{\nu} \chi$ diverge:
\begin{equation}
\nabla_{z}\nabla_{z} \chi \sim \frac{\epsilon}{z} \chi_{,z}
\end{equation}
\begin{equation}
\nabla_{\bar{z}} \nabla_{\bar{z}} \chi \sim \frac{\epsilon}{\bar{z}} \chi_{,\bar{z}}
\end{equation}
Consider first the $zz$ component of the gravity equation. Term by term, we find the following contributions to the $\epsilon/z$ divergence:
\begin{equation}\label{div1}
G_{zz} \sim -\frac{\epsilon}{z}K_{z}
\end{equation}
\begin{equation}\label{div2}
\frac{1}{2}\gamma \partial^{\rho} \chi \partial^{\sigma} \chi R_{z \rho z \sigma} \sim -\frac{\epsilon}{2z}\gamma K_{z}^{ij} \chi_{0,i} \chi_{0,j}
\end{equation}
\begin{equation}\label{div3}
\frac{\gamma}{2} (\nabla_{z} \nabla^{\rho} \chi)(\nabla_{z} \nabla_{\rho} \chi) \sim 2  \frac{\epsilon}{z} \gamma  \chi_{z} \chi_{z\bar{z}}
\end{equation}
\begin{equation}\label{div4}
-\frac{\gamma}{2} (\nabla_{z}\nabla_{z} \chi) \Box \chi \sim - \frac{\gamma}{2} \frac{\epsilon}{z} \chi_{z} (4 \chi_{z\bar{z}} + h^{ij} D_{i} D_{j}\chi_{0})
\end{equation}
\begin{equation}\label{div5}
-\frac{\gamma}{2} \frac{1}{2} G_{zz} (\partial \chi)^{2}  = \frac{\gamma \epsilon}{4z} K_{z} h^{ij} \chi_{0,i} \chi_{0,j}
\end{equation}
and all other terms in the equation of motion are non-singular.\\
Let us sketch out how to obtain some of the expressions above in details. In most of the equations above (namely, equations (\ref{div2}), (\ref{div4}) and (\ref{div5})), the $(\epsilon/z)$ factor comes from the metric (through the curvature tensor or the Christoffel symbols), and for the scalar part it suffices to use the scalar field expansion at $\epsilon=0$ (equation (\ref{chiRNCepsilon0})) \footnote{More properly, suppose we insist on using the expansion (\ref{chiRNC}) at finite $\epsilon$. Then equations (\ref{div2}), (\ref{div4}) and (\ref{div5}) each will take the form of the product of $(\epsilon/z)$ and a Taylor series in $(z\bar{z})^{\epsilon}$. In other words, we have a hierarchy of divergences. This is a manifestation of the ``splitting problem'' on the level of the divergences of the equation of motion.\\
However, we are trying to demonstrate that the coefficient of the $(\epsilon/z)$ yields the minimal surface as $\epsilon \rightarrow 0$. Thus we have simply evaluated the coefficient of $(\epsilon/z)$ at $\epsilon=0$, which collapses the hierarchy and amounts to plugging in the scalar field expansion at zero $epsilon$ at the outset.}.\\
In the case of equation (\ref{div3}), however, the divergence coming from the metric is actually slightly sublinear, but the scalar expansion (\ref{chiRNC}) for $\epsilon \neq 0$ actually results in a slight enhancement of the divergence from sublinear to linear. To see this, note that when we expand the sum over $\rho$ in (\ref{div3}), the only singular term is $2 g^{z\bar{z}} (\nabla_{z}\nabla_{z}\chi) (\nabla_{z} \nabla_{\bar{z}} \chi)$. Expanding the second covariant derivatives into partial derivatives and Christoffel symbols, we further find that the leading divergent part of this expression is $4 \rho^{2\epsilon}\frac{\epsilon}{z}\chi_{,z}\chi_{,z\bar{z}}$, which is slightly subleading compared to the $\epsilon/z$ divergence. Using (\ref{chiRNC}), however, we find that $\chi_{,z\bar{z}}$ secretly contains a factor of $\rho^{-2\epsilon}$:
\begin{equation}
\chi_{,z\bar{z}} = \chi_{z\bar{z}}(1-\epsilon)^{2} (z\bar{z})^{-\epsilon} + \dots
\end{equation}
Thus, there is an enhancement from a slightly subleading divergence (of the order of $\rho^{2\epsilon}\frac{\epsilon}{z}$ to a leading divergence ($\frac{\epsilon}{z}$), and we obtain (\ref{div3}). Note also the delicate cancellation of the terms with $\chi_{z\bar{z}}$ in (\ref{div3}) and (\ref{div4}).\\
Now we add up (\ref{div1})-(\ref{div5}) and demanding that the $\epsilon/z$ divergence cancels. We then find the condition:
\begin{equation}\label{zzdiv}
 \left( -1 + \frac{\gamma G}{4} h^{ij} \chi_{,i} \chi_{,j} \right) K_{z} - \frac{\gamma G}{2} K_{z}^{ij} \chi_{,i}\chi_{,j} - \frac{\gamma G}{2} \chi_{,z} h^{ij} D_{i}D_{j}\chi = 0
\end{equation}
where we dropped the subscript $0$ in $\chi_{0}$ (and replaced it simply by $\chi$), and also replaced $\chi_{z}$ by $\chi_{,z}$ it is now implicitly understood that this is an equation at $\epsilon=0$ and at $z=\bar{z}=0$. Comparing with the surface equation (\ref{SurfaceEqn}), we see that this is precisely the same equation. Thus, the Lewkowycz-Maldacena prescription of fixing the surface by the divergence of the equation of motion works at least with respect to the $1/z$ divergence of the $zz$ component of the equation of motion.\\
We end this section by mentioning the divergences appearing in the other components of the gravity equation of motion as well as the scalar equation. The divergence of the $\bar{z}\bar{z}$ component is the same as that of the $zz$ component, except for the substitution $z \rightarrow \bar{z}$. The $zi$ component has no first order divergence, but it does have slightly subleading ones (of order $z^{1-\epsilon}$).\\
As for the $z\bar{z}$ component, two of the terms actually have a quadratic divergence:
\begin{equation}
\frac{\gamma}{2} (\nabla_{z}\nabla^{\rho}\chi) (\nabla_{\bar{z}}\nabla_{\rho}\chi) \sim \gamma e^{-2A} \frac{\epsilon^{2}}{z\bar{z}} \chi_{z}\chi_{\bar{z}} + \dots
\end{equation}
\begin{equation}
-\frac{\gamma}{4}g_{z\bar{z}} (\nabla^{\rho}\nabla^{\sigma} \chi) (\nabla_{\rho}\nabla_{\sigma}\chi) \sim -\gamma e^{-2A} \frac{\epsilon^{2}}{z\bar{z}} \chi_{z} \chi_{\bar{z}} + \dots
\end{equation}
where $\dots$ stands for subleading divergences (including linear). But the quadratic divergences in these two terms exactly cancel out each other !\\
Finally, we consider the $ij$ component. This one has a genuine quadratic divergence:
\begin{equation}
    \frac{\gamma}{4}h_{ij} (\nabla^{\rho}\nabla^{\sigma}\chi)(\nabla_{\rho}\nabla_{\sigma}\chi) \sim -2\gamma h_{ij} e^{-4A} \frac{\epsilon^{2}}{z\bar{z}} \chi_{z} \chi_{\bar{z}}
\end{equation}
Similar, the equation of motion for the scalar field also has a quadratic divergence:
\begin{equation}
-\gamma G^{\mu\nu} \nabla_{\mu}\nabla_{\nu}\chi \sim \gamma \frac{8\epsilon^{2}}{z\bar{z}} e^{-4A} (K_{\bar{z}}\chi_{z} + K_{z}\chi_{\bar{z}})
\end{equation}
The presence of the subleading divergences as well as quadratic divergences is somewhat troublesome, but it is a feature that Horndeski theory shares with higher derivative/higher curvature gravity (see for example \cite{Bhattacharyya:2014yga}). In fact, the divergence structure is strikingly similar: the paper \cite{Bhattacharyya:2014yga} shows that the $ij$ component of the equation of motion also suffers from a quadratic divergence in Gauss-Bonnet gravity. We will leave the question of how to get rid of these other divergences to future work, but the work \cite{Camps:2014voa} is a promising step in this direction: the authors of \cite{Camps:2014voa} show that an ansatz more general than (\ref{RNC}) is needed to cancel the subleading divergences. We will come back to this point in the Conclusion section.

\section{Comments on the thermal entropy}\label{Sec:ThermalS}
In this section, we revisit the issue of the thermal entropy for the black hole solution in Section \ref{Sec:Numerics}. As pointed out in the literature \cite{Feng:2015oea}, the standard methods of deriving the entropy (Wald's entropy formula, the Iyer-Wald formalism, and the Euclidean method) seem to yield conflicting answers. We will make the case that the correct entropy should be the one given in (\ref{SthermalHorndeski}), i.e. the Wald entropy (which happens to coincide with the usual Bekenstein-Hawking entropy).\\
As shown in \cite{Feng:2015oea}, Horndeski gravity admits black hole solutions given by: 
\begin{equation}\label{StationaryBHAnsatz}
ds^{2} = -h{(r)}dt^{2} + \frac{dr^{2}}{f{(r)}} + r^{2} d\Omega_{d-2,\epsilon}^{2}
\end{equation}
\begin{eqnarray}
h{(r)} &=& \frac{(d-1)^{2}\beta^{2}\gamma^{2}g^{4}r^{4}}{\epsilon(d+1)(d-3)(4\kappa+\beta\gamma)^{2}} {}_{2}F_{1}{\left(1,\frac{1}{2}(d+1),\frac{1}{2}(d+3),-\frac{d-1}{(d-3)\epsilon}g^{2}r^{2}\right)} \nonumber \\
&-& \frac{\mu}{r^{d-3}} + \frac{8\kappa[g^{2}r^{2}(2\kappa+\beta\gamma)+2\epsilon\kappa]}{(4\kappa+\beta\gamma)^{2}}
\end{eqnarray}
\begin{equation}
f{(r)} = \frac{(4\kappa+\beta\gamma)^{2} [(d-1)g^{2}r^{2}+(d-3)\epsilon]^{2}}{[(d-1)(4\kappa+\beta\gamma)g^{2}r^{2} + 4(d-3)\epsilon\kappa ]^{2}} h{(r)}
\end{equation}
\begin{equation}\label{dchidr}
\left( \frac{d\chi}{dr} \right)^{2} = \frac{\beta}{f} \left( 1 + \frac{(d-3)\epsilon}{(d-1)g^{2}r^{2}} \right)^{-1}
\end{equation}
where $d$ is the dimension, and $\epsilon = -1,0,1$ correspond to a hyperbolic, planar and spherical horizon, respectively. Also, the constants $g$ and $\beta$ are related to the couplings $\alpha$ and $\Lambda$ in the action by:
\begin{equation}\label{alphaRelation}
\alpha = \frac{1}{2}(d-1)(d-2)g^{2}\gamma
\end{equation}
\begin{equation}\label{LambdaRelation}
\Lambda = -\frac{1}{2}(d-1)(d-2)g^{2} \left( 1+\frac{\beta\gamma}{2\kappa} \right)
\end{equation}

\subsection{Black hole entropy from Iyer-Wald formalism}
Following \cite{Feng:2015oea}, let us review the computation of black hole entropy used the Iyer-Wald formalism. This formalism gives us 2 statements related to the entropy: (1) the first law of black hole mechanics, and (2) the statement relating the integral of the Noether charge (of diffeomorphism invariance) over the bifurcation surface of the black hole to the entropy.\\
To obtain the first statement (the first law), we compute a closed differential form $\delta Q - \xi \cdot \Theta$, where $\delta Q$ is an on-shell perturbation of the Noether charge (of diffeomorphism invariance), $\xi$ is the bifurcate timelike Killing vector field of the black hole, and $\Theta$ is the boundary term of the gravity action. The first law of black hole thermodynamics then comes out as the equation:
\begin{equation}\label{IyerWald2}
\int_{\infty} \delta Q - \xi \cdot \Theta = \int_{\mathcal{H}} \delta Q - \xi \cdot \Theta
\end{equation}
where, on the left hand side, we integrate this differential form on a sphere at infinity, and on the right-hand side we integrate on the bifurcation surface of the black hole. In the case of Einstein gravity, the left-hand side above coincides with the mass perturbation $\delta M$, and the right-hand side coincides with $T\delta S$.\\
The second statement of the Iyer-Wald formalism tells us that:
\begin{equation}\label{IyerWald1}
\int_{\mathcal{H}} Q = TS
\end{equation}
In other words, the integral of $Q$ over the bifurcation surface equals the product of the temperature and the entropy.\\
Let us now examine what happens to these 2 statements in the case of Horndeski gravity. The Noether charge for a stationary black hole metric of the form (\ref{StationaryBHAnsatz}) is given in \cite{Feng:2015oea}. If we write it as $Q = Q_{Einstein} + Q_{\gamma}$ where $Q_{Einstein}$ is the contribution from Einstein gravity and the minimal coupling, and $Q_{\gamma}$ is the contribution from the non-minimal coupling, we have:
\begin{equation}
Q_{Einstein} = \frac{r^{d-2}}{16\pi G} \sqrt{\frac{f}{h}} h' \Omega_{d-2}
\end{equation}
\begin{equation}
Q_{\gamma} = -\frac{1}{32\pi}(d-2)\gamma r^{d-3} \sqrt{\frac{h}{f}}f^{2} \chi'^{2} \Omega_{d-2}
\end{equation}
From the above, it is easily seen that $Q_{\gamma}$ vanishes on the horizon. To see this, we use (\ref{dchidr}) to substitute for $(\chi')^{2}$, and note the fact that $h/f$ is regular on the horizon. Thus the integral of $Q$ on the bifurcation surface reduces to that of $Q_{Einstein}$, and equals the product $TS$ with $S$ given by the Wald entropy !\\
The fact that the identity (\ref{IyerWald1}) is consistent with the Wald entropy, i.e. consistent with the squashed cone method, should not be surprising. Indeed, there exist quite rigorous arguments (see for example \cite{Nelson:1994na, Iyer:1995kg}) that the black hole entropy as derived by the conical singularity method always coincides with the entropy obtained from the integral of the Noether charge.\\
Things do go wrong for the first law, however: it can be seen from (\ref{dchidr}) that the derivative of the scalar field in the radial direction diverges at the horizon, and the scalar field itself has a branch cut singularity there. As the analysis in \cite{Feng:2015oea} shows, what happens is that the variational identity (\ref{IyerWald2}) continues to hold (since it follows from Stokes theorem), but the two sides of this equation can no longer be identified as $\delta M$ and $T\delta S$ (with $S$ taken to be the Wald entropy). Explicitly, we obtain \cite{Feng:2015oea} for the planar case ($\epsilon=0$):
\begin{equation}\label{chiInfinityPlanarBH}
\int_{\infty} \chi = \frac{(d-2)}{16\pi G} \left( 1 + \frac{\gamma\beta G}{4} \right) \delta \mu
\end{equation}
\begin{equation}\label{chiHorizonPlanarBH}
\int_{\mathcal{H}} \chi = \frac{(d-1)(d-2)g^{2}}{16\pi G} \left( 1 + \frac{\gamma\beta G}{4} \right) r_{0}^{d-2} \delta r_{0}
\end{equation}
where the terms contaning $\gamma$ above essentially arise from the singularity of the scalar field mentioned above. In view of this difficulty, the authors of \cite{Feng:2015oea} proceed by simply \textit{defining} the left-hand side of (\ref{IyerWald2}) as $\delta M$ and the right-hand side as $T\delta S$. From (\ref{chiInfinityPlanarBH}) and (\ref{chiHorizonPlanarBH}), the authors of \cite{Feng:2015oea} then obtain the following \textit{definitions} for the mass and the entropy:
\begin{equation}\label{MPope}
\tilde{M} = \frac{(n-2)}{16\pi G} \left( 1 + \frac{\gamma \beta G}{4} \right) \mu
\end{equation}
\begin{equation}\label{SPope}
\tilde{S} = \frac{1}{4G} \left( 1 + \frac{\gamma \beta G}{4} \right) r_{+}^{n-2}
\end{equation}
While the definitions above for $M$ and $S$ have the virtue that the first law is automatically satisfied, it is important to keep in mind that they are merely definitions. In the case of Einstein gravity, there exist indepdendent, nontrivial checks for the mass and the entropy: the mass in that case coincides with the Komar integral, and the entropy is of course the Bekenstein-Hawking entropy, which obeys the second law for example. In the case of Horndeski gravity, there are no independent checks of (\ref{MPope}) and (\ref{SPope}).\\
On the other hand, we have seen that from the viewpoint of holographic entanglement, it is much more natural to take the thermal entropy to be the usual Wald entropy since this is what we obtain from entanglement entropy as the size of the boundary region approaches the whole boundary. To summarize, for planar black holes we have the following mass and thermal entropy:
\begin{equation}\label{MWald}
M = \frac{(n-2)}{16\pi G} \mu \qquad\qquad\qquad S = \frac{1}{4G} r_{+}^{n-2}.
\end{equation}
The corresponding expressions for $\epsilon=1$ are somewhat more complicated, and can be found in \cite{Feng:2015oea}.\\

\subsection{Black hole entropy from conical singularity}
Even though the analysis above should assure us that the Wald entropy is correct, there is potentially a loophole because the singular behavior of the scalar field on the horizon contradicts the assumption of regularity of the scalar field used in the derivation of the entanglement entropy functional. Indeed, near the horizon, the scalar field expands as:
\begin{equation}\label{NearHorizon}
    \chi = \chi_{0} + \frac{2\sqrt{(d-1)\beta}gr_{0}^{3/2}}{(d-1)g^{2}r_{0}^{2}+(d-3)\epsilon} \sqrt{r-r_{0}} + \mathcal{O}{(r-r_{0})^{3/2}}
\end{equation}
Thus, the assumption of regularity used to derive the expansion (\ref{chiRNC}) technically does not apply in the case of the thermal entropy. In this subsection, we take a closer look at this and argue that - despite this singularity of the scalar field on the horizon - the conical singularity method should still yield the Wald entropy.\\
First, we need to redefine the radial coordinate from the Schwarzschild-like $r$ to the coordinate $\rho$ used in (\ref{chiRNC}) to facilitate comparison between the two near-horizon expansions. Note that the coordinate $\rho$ in (\ref{chiRNC}) satisfies two properties: the horizon is at $\rho=0$, and the near-horizon metric looks like flat space in polar coordinates: $ds^{2} = d\rho^{2} + \rho^{2} d\tau^{2}$ when we turn off the conical singularity ($\epsilon = 0$). Note also that the usual steps taken to derive the Hawking temperature of a black hole involves precisely a coordinate redefinition with the two properties above, and this is the procedure we will follow to obtain the transformation from $r$ to $\rho$. Near the horizon, the functions $f(r)$ and $h(r)$ in the metric expand as:
\begin{equation}
    f{(r)} = f_{1}(r-r_{0}) +\mathcal{O}{(r-r_{0})^{2}}
\end{equation}
\begin{equation}
    h{(r)} = h_{1}(r-r_{0}) + \mathcal{O}{(r-r_{0})^{2}}
\end{equation}
for some coefficients $f_{1}$ and $h_{1}$. From the above we obtain the desired coordinate redefinition:
\begin{equation}
    \rho = \frac{2}{\sqrt{f_{1}}} \sqrt{r-r_{0}}
\end{equation}
The near horizon expansion (\ref{NearHorizon}) in terms of $\rho$ now reads:
\begin{equation}\label{NearHorizonRho}
    \chi = \chi_{0} + \chi_{1}\rho + \dots
\end{equation}
for some coefficient $\chi_{1}$. The expansion above is supposed to replace the expansion (\ref{chiRNC}) at $\epsilon=0$, so let us compare the two. The expansion above is a function of $\rho$ alone, whereas (\ref{chiRNC}) at $\epsilon=0$ allows for angular dependence. This is expected due to the $U(1)$ symmetry of the black hole (i.e. time translation symmetry) which is not present in (\ref{chiRNC}). Secondly, the leading power of $\rho$ is first order in the expansion above, whereas it is quadratic in (\ref{chiRNC}) at $\epsilon = 0$. This is, of course, due to the fact that one expansion is regular near $\rho=0$ and the other is not. We also remark that, in terms of $\rho$, the singularity of the scalar field is only a ``kink'' near $\rho=0$ as opposed to a divergence (\ref{NearHorizon}). That the nature of this singularity depends on the coordinate used should not be surprising; moreover it was noted in \cite{Feng:2015oea} already that the singularity is milder than it seems, in the sense that coordinate-invariant quantities do not suffer from divergences across the horizon.\\
The challenge now is to figure out how expansion (\ref{NearHorizonRho}) is affected when we turn on $\epsilon$, because we need the expansion at finite $\epsilon$ to derive entanglement entropy. Equivalently, we need to find out the leading power of $\tilde{\rho}$ in the parent space. We remark that this is best done by first computing the parent-space analog of the Horndeski black hole, then go to the quotient space and expanding near the horizon. This is clearly a very nontrivial task in general, and few exact solutions of this type are known (however an explicit parent-space analog of the hyperbolic black hole is known \cite{Emparan:1999gf}). Fortunately, we can deduce the powers of $\tilde{\rho}$ in the parent space based on (\ref{NearHorizonRho}): the smallest power of $\tilde{\rho}$ consistent with a first power in $\rho$ at $\epsilon=0$ is a first power in $\tilde{\rho}$. Thus, in the parent space:
\begin{equation}
    \chi  = \chi_{0} + \tilde{\chi}_{1} \tilde{\rho} + \dots
\end{equation}
In the quotient space, this translates to:
\begin{equation}\label{QuotientBH}
    \chi = \chi_{0} + \chi_{1} \rho^{1-\epsilon} + \dots
\end{equation}
which, at $\epsilon=0$, reduces to (\ref{NearHorizonRho}). Of course, there are infinitely many higher powers of $\tilde{\rho}$ which are also consistent with first order in $\rho$ at $\epsilon=0$. For example, $\tilde{\rho}^{n}$ gives $\rho$ (with no $\epsilon$ dependence) etc.\\
Can the anomalous power of $\chi$ in (\ref{QuotientBH}) result in a new contribution to the black hole entropy ? It is true that one can extract a term first order in $\epsilon$ from the expansion above: $\delta_{\epsilon} \chi = -\epsilon \chi_{1} \rho \log{\rho}$. However, the fact that $\rho \log{\rho}$ vanishes as $\rho \rightarrow 0$ means this term cannot give rise to any new contribution when we plug it into the action and integrate over a small region near the tip of the cone. For example, the minimal coupling term gives:
\begin{equation}
    \int_{\rho \approx 0} \rho d\rho d\tau \delta_{\epsilon} \left( g^{\mu\nu} \chi_{,\mu} \chi_{,\nu} \right) \propto \int_{\rho \approx 0} \rho d\rho \chi_{\rho} (\delta_{\epsilon} \chi)_{,\rho} = 0
\end{equation}
similarly for the non-minimally coupled term.\\
To summarize, we think that the correct entropy should be the Wald entropy for 3 reasons: (1) the Wald entropy is consistent with the Iyer-Wald formalism, in the sense that it is consistent with the integral of $Q$ over the horizon; (2) a rederivation of the black hole entropy from the conical method does not seem to yield any new contribution on top of the Wald entropy, and (3) the Wald entropy coincides with the limit of entanglement entropy as the boundary region approaches the whole boundary.
 
\section{Numerical applications}\label{Sec:Numerics}
In this section, we find numerically the entanglement entropy of black hole solutions of Horndeski gravity given in \cite{Feng:2015oea}. 

\subsection{3-dimensional planar black hole}
\begin{figure}[!t]
\centering
\begin{subfigure}{.4\textwidth}
\includegraphics[width=\textwidth]{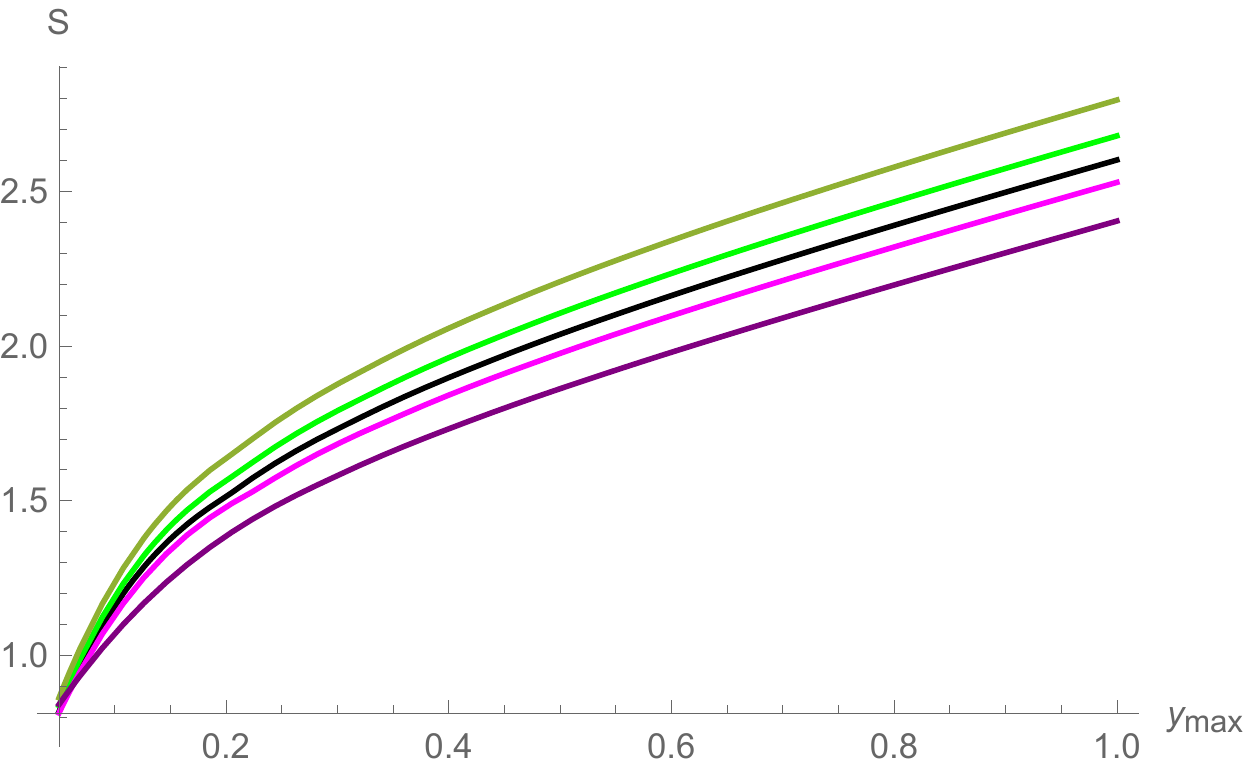}
\caption{}
\label{Fig:Concavity}
\end{subfigure}
\qquad\qquad
\begin{subfigure}{.4\textwidth}
\includegraphics[width=\textwidth]{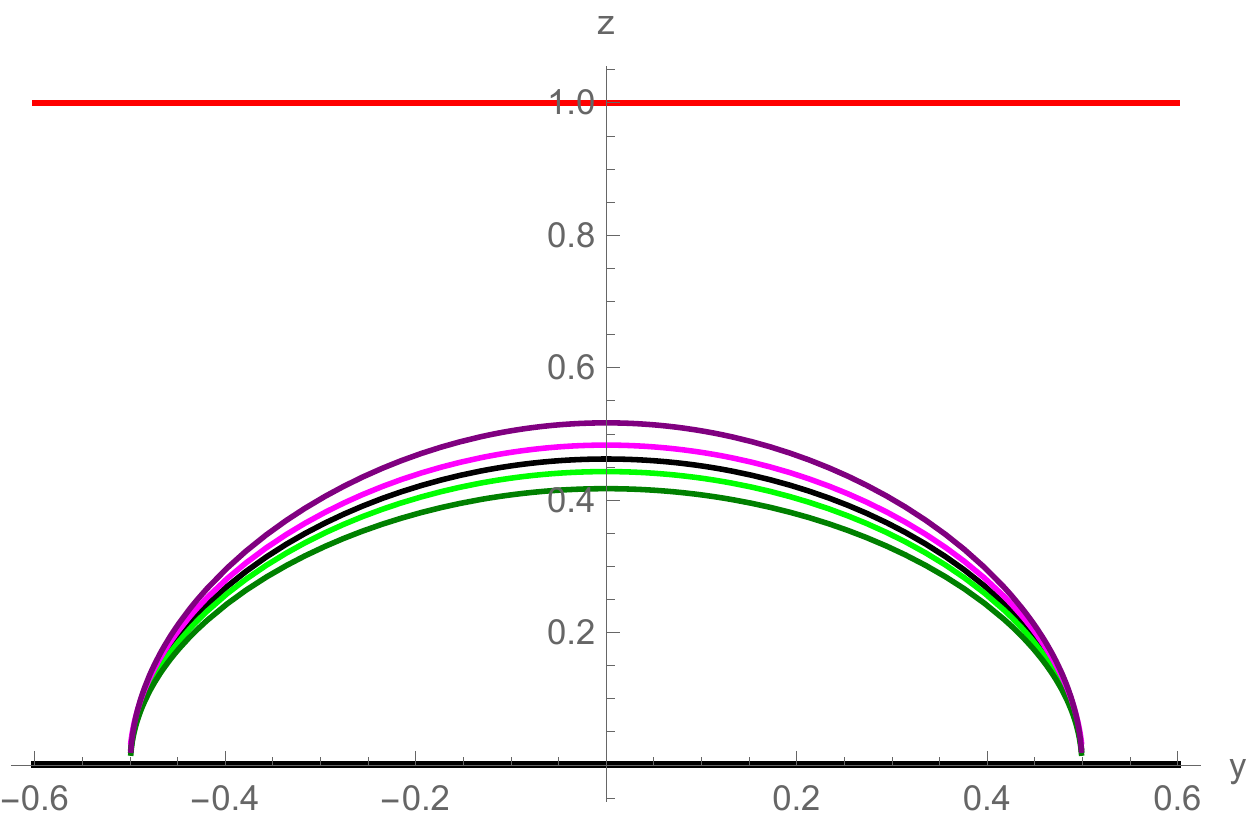}
\caption{}
\label{Fig:PlanarBHSurfaces}
\end{subfigure}
\caption{$d=3$ planar black hole. a)Entanglement entropy as a function of the half-width  of the boundary interval.  b) Some representative minimal surfaces. In both plots we have used $g=\mu=\beta=G=1$, and the values of $\gamma$ are: $\gamma=0$ (black), $\gamma=-0.05$ (light green), $\gamma=-0.125$ (dark green), $\gamma=0.05$ (light purple) and $\gamma=0.125$ (dark purple).}
\label{fig:planar_sfces_concav}
\end{figure}

Let us first specialize to the 3-dimensional, planar black hole ($d=3$, $\epsilon=0$). The metric and the scalar field profile simplify to:
\begin{equation}
ds^{2} = -(g^{2}r^{2}-\mu)dt^{2} + \frac{dr^{2}}{g^{2}r^{2}-\mu} + r^{2}dx^{2}
\end{equation}
\begin{equation}
\chi = \frac{\sqrt{\beta}}{g} \log{(gr + \sqrt{(gr)^{2}-\mu})} + \chi_{0}
\end{equation} 
Note that the metric is exactly the BTZ black hole. The horizon is located at $r_{+} = \sqrt{\mu}/g$. We will find it convenient to go to the Fefferman-Graham coordinate:
\begin{equation}
\zeta = \frac{g}{gr+\sqrt{g^{2}r^{2}-\mu}}
\end{equation}
and rescale the boundary coordinates as $\tau = t/2$ and $y =x/(2g)$, the metric then becomes:
\begin{equation}
ds^{2} = \frac{1}{g^{2} \zeta^{2}} \left[ - (g^{2}-\mu \zeta^{2})^{2} d\tau^{2} + d\zeta^{2} + (g^{2}+\mu \zeta^{2})^{2} dy^{2} \right]
\end{equation}
and the scalar field in terms of $z$ satisfies:
\begin{equation}
\frac{d\chi}{d\zeta} = -\frac{\sqrt{\beta}}{g\zeta}
\end{equation}
Let us now parametrize the Ryu-Takayanagi surface as $X^{\mu} = (\tau=\mathrm{const},y,\zeta(y))$. The functional to be minimized then takes the form:
\begin{equation}
S_{EE} = \int dx \frac{\sqrt{(\zeta')^{2}+(g^{2}+\mu \zeta^{2})^{2}}}{g\zeta} \left( 1 + \tilde{\gamma} \frac{(\zeta')^{2}}{(\zeta')^{2}+(g^{2} + \mu \zeta^{2})^{2}} \right)
\end{equation}
where we defined $\tilde{\gamma} = \frac{\gamma G \beta}{4}$. We minimized the functional above numerically and plot in Figure \ref{Fig:Concavity} the entanglement entropy versus the half-width $y_{max}$ of the boundary interval (which ranges from $-y_{max}$ to $y_{max}$). For completeness, we present 3 cases: the Einstein gravity case $\gamma = 0$, a case with $\gamma > 0$ and a case with $\gamma < 0$, even though from the viewpoint of bulk causality $\gamma$ is required to be non-positive \cite{Minamitsuji:2015nca}. In Figure \ref{Fig:PlanarBHSurfaces}, we present the plot of a few RT surfaces for a few different values of $\gamma$.\\

Let us comment that all three curves in Figure \ref{Fig:Concavity} are concave. Concavity is one of the hallmark features of entanglement entropy (in fact, of entropy of any kind), and implies the property of strong subadditivity. From the holographic viewpoint, concavity can be expected from the fact that the holographic entanglement entropy is the minimization of a functional. Indeed, the proof of strong subadditivity in the usual Einstein gravity case can be generalized in a straightforward way to any extensive functional \cite{Headrick:2007km}.

\subsection{3-dimensional, spherical black hole}
\begin{figure}[!t]
\centering
\begin{subfigure}[b]{0.4\textwidth}
\includegraphics[width=\textwidth]{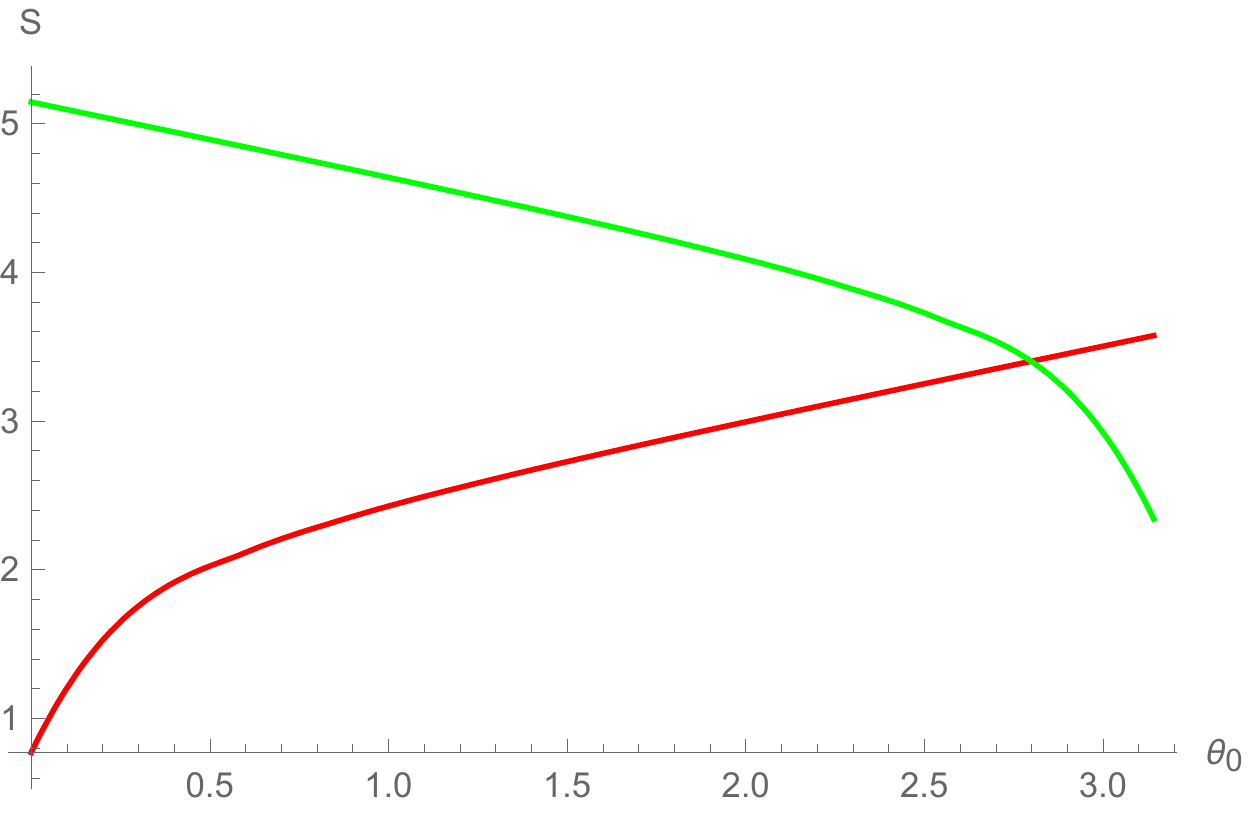}
\caption{}
\label{Fig:PhaseTransition}
\end{subfigure}
\qquad\qquad
\begin{subfigure}[b]{0.4\textwidth}
\includegraphics[width=\textwidth]{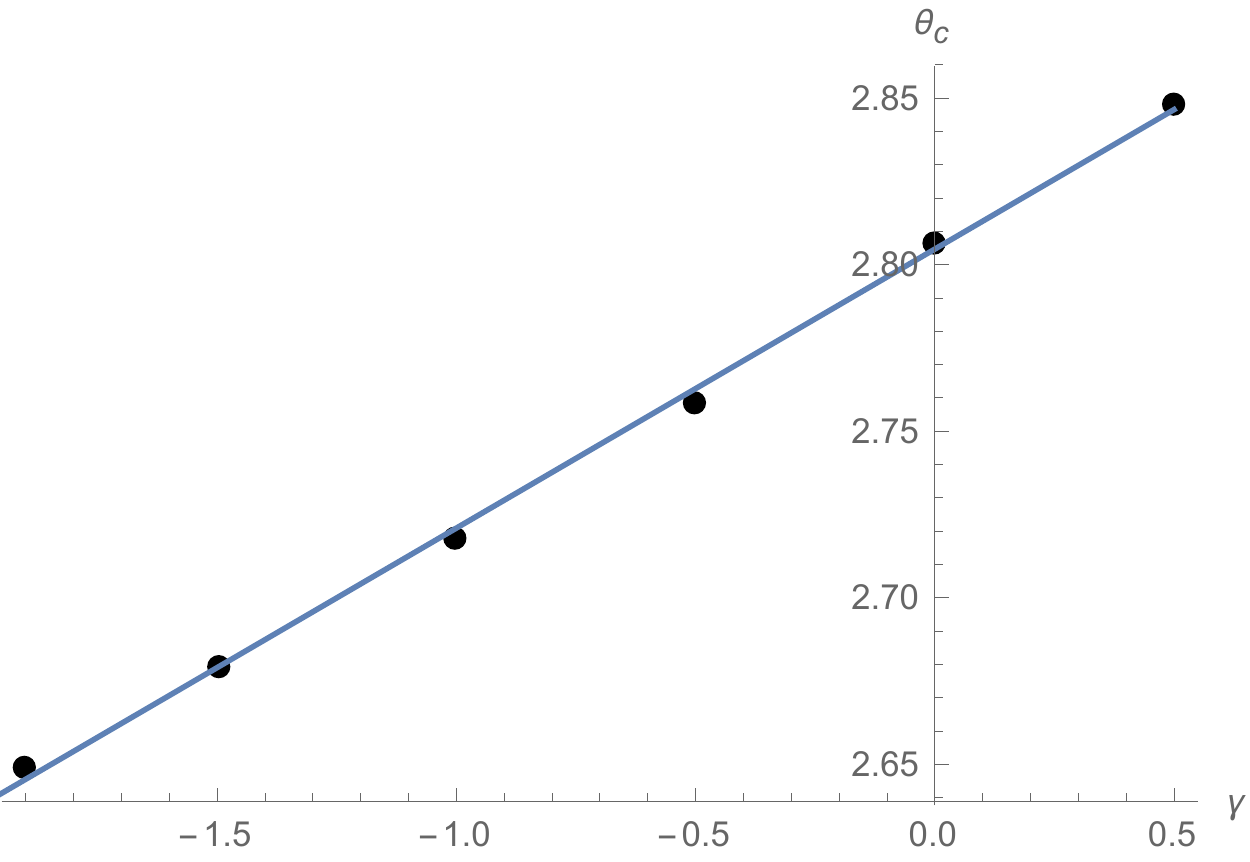}
\caption{}
\label{Fig:CritAngleVsGamma}
\end{subfigure}
\caption{$d=3$. a) shows the Wald entropy for the connected (blue) and disconnected (orange) surfaces as a function of the boundary region $\theta_{0}$. We have used $\mu'=g=\beta=G=1$ and $\gamma=-0.1$. b)  Critical angle $\theta_c$ versus $\gamma$, also with $g=\mu'=\beta=G=1$.   }
\end{figure}

Next, we consider the 3-dimensional spherical black hole ($d=3,\epsilon=1$). The metric is still the BTZ metric:
\begin{equation}
ds^2=-\left(-\mu^\prime +g^2r^2\right)dt^2+\frac{dr^2}{\left(-\mu^\prime +g^2r^2\right)}+r^2d\phi^2
\end{equation}
where
\begin{equation}
\mu^\prime=\mu-\frac{16\kappa^2}{(\beta\gamma+4\kappa)^2}
\end{equation}
and the scalar field profile is identical to that of the 3-dimensional planar black hole.\\
As usual, for a given boundary interval of half-width $\theta_{0}$ there are two minimal surfaces satisfying the homology constraint: a connected one and a disconnected one which includes the horizon as a connected component. Intuition from Einstein gravity suggests that there exists a phase transition from the connected one to the disconnected as we increase the size $\theta_{0}$ of the boundary region.\\
In Figure \ref{Fig:PhaseTransition}, we vary the size $\theta_{0}$ of the boundary interval from $0$ to $\pi$ and plot the value of the functional on evaluated on the connected surface and disconnected surface. As expected, the curve for the connected surface increases monotonically while the the curve for the disconnected one decreases with increasing $\theta_{0}$. The two curves intersect at an angle $\theta_{c}$ (around 2.8 rad), at which point the phase transition happens (since the disconnected one yields a smaller value of the functional beyond this angle, and the homology constraint instructs us to use whichever surface has the smaller value).\\
Like in Einstein gravity, this phase transition has an interesting consequence for the Araki-Lieb inequality, which tells us that the difference between the entropy of a subregion $A$ and the entropy of its complement is at most the entropy of the mixed state (the thermal entropy in this case):
\begin{equation}
|S_{A} - S_{A^{C}}| \leq S_{\mathrm{thermal}}
\end{equation}
For $\theta_{0} \geq \theta_{c}$, the phase transition implies that the difference above is exactly equal to the thermal entropy, hence the Araki-Lieb inequality is saturated. Put differently, if we were to plot $|S_{A} - S_{A^{C}}|/S_{\mathrm{thermal}}$ as a function of $\theta_{0}$, we would see an \textit{entanglement plateau} for $\theta_{0} > \theta_{c}$.\\
Next, we keep the values of the constants $g$, $\mu$ and $\beta$ fixed (to unity) and vary the value of the coupling $\gamma$. For each such $\gamma$ we computed the angle $\theta_{c}$ of the phase transition, and we plot in Figure \ref{Fig:CritAngleVsGamma} the critical angle as a function of $\gamma$. In this plot we focus on the negative $\gamma$ regime, since this is the regime consistent with bulk causality (see for instance \cite{Minamitsuji:2015nca}).
Note that keeping $g$, $\mu$ and $\beta$ fixed means both the metric and the scalar field profile are kept fixed as we vary $\gamma$. However, because of the relations (\ref{alphaRelation}) and (\ref{LambdaRelation}), this means the value of $\alpha$ and $\Lambda$ are actually varied, so that the different curves belong to different Horndeski theories. Also, for $\gamma = 0$, Figure \ref{Fig:CritAngleVsGamma} is consistent with the analytical value of the critical angle which was computed analytically in \cite{Hubeny:2013gta} for the usual BTZ black hole:
\begin{equation}
\theta_{c} = \frac{1}{r_{+}} \mathrm{arccoth}{(2\mathrm{coth}{(\pi r_{+})})-1}
\end{equation}
In terms of the entanglement plateau, this means that the plateau becomes larger and larger as we make $\gamma$ more and more negative.

\subsection{4-dimensional, spherical black hole}
\begin{figure}[!t]
\centering
\begin{subfigure}[b]{0.4\textwidth}
\includegraphics[width=\textwidth]{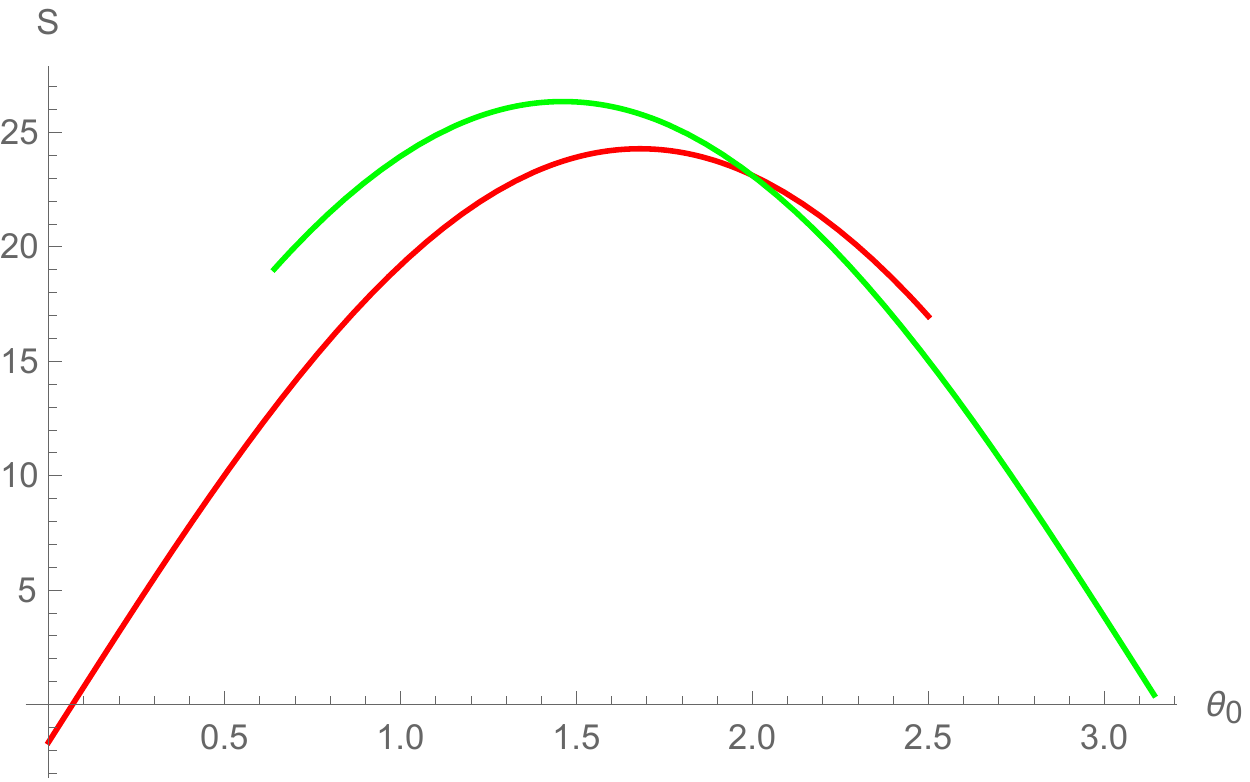}
\caption{}
\label{Fig:PhaseTransition4d}
\end{subfigure}
\qquad\qquad
\begin{subfigure}[b]{0.4\textwidth}
\includegraphics[width=\textwidth]{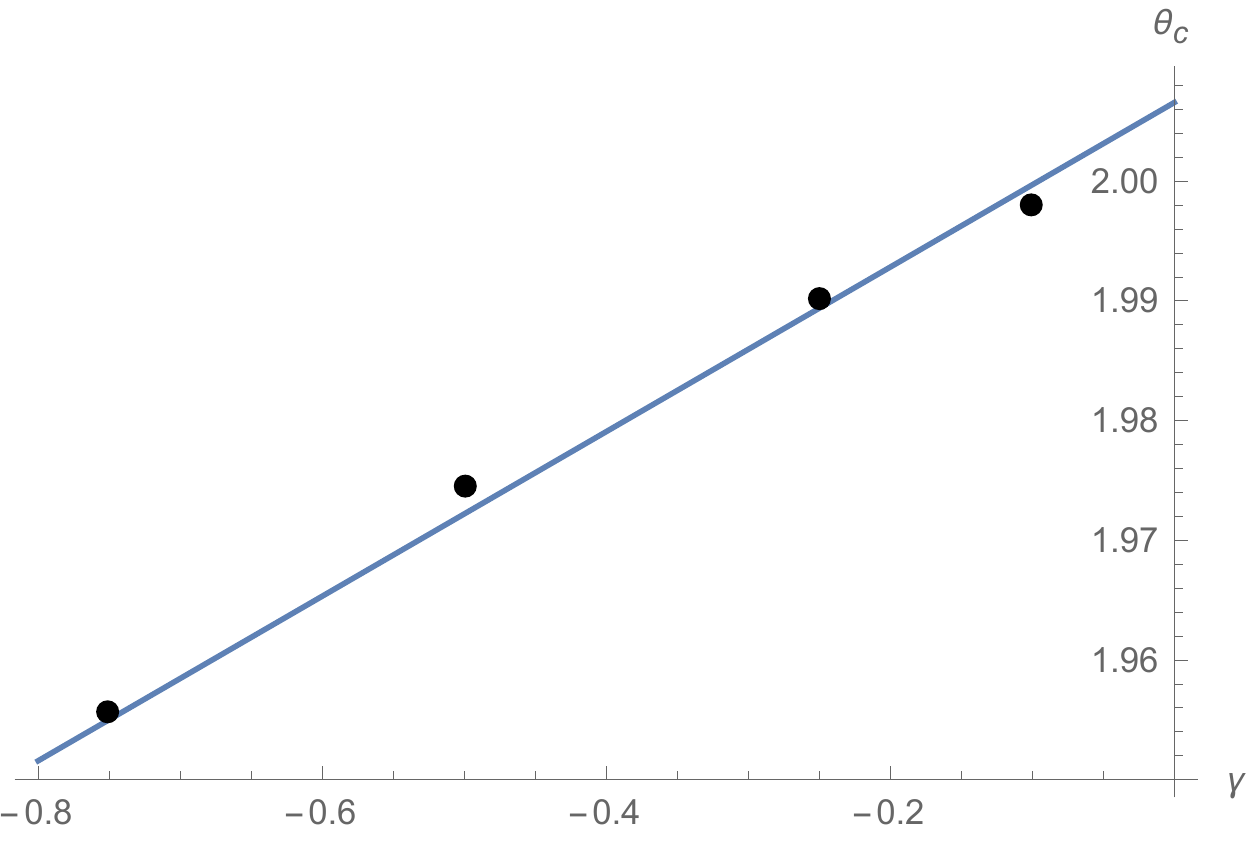}
\caption{}
\label{Fig:CritAngleVsGamma4d}
\end{subfigure}
\caption{$d=4$. a) shows the  Wald entropy for the connected (red) and disconnected (green)  surfaces as a function of the boundary region $\theta_{0}$. We have used $g=\beta=G=1$, $\gamma=-0.1$ and $\mu = 2.711055$. b)  Critical angle $\theta_c$ versus $\gamma$, with $g=\mu=\beta=1$.   }
\end{figure}

In this subsection, we present the phase transition for a higher-dimensional case: the 4-dimensional black hole. We relegate the plots of the minimal surfaces themselves to Appendix \ref{App:RTSurfaces2}. From this Appendix, a noteworthy feature of the RT surfaces is that the connected surface stops existing for sufficiently large $\theta_{0}$. Equivalently, the disconnected surface does not exist for sufficiently small $\theta_{0}$. This feature is potentially worrisome, since it implies that the competition between the two kinds of surfaces only exist within a range of $\theta_{0}$ smaller than $(0,\pi)$. The phase transition, thus, must occur within this band.\\
In Figure \ref{Fig:CritAngleVsGamma4d}, we plot the angle $\theta_{c}$ of the phase transition as a function of the non-minimal coupling $\gamma$ at fixed $g$ (which corresponds to the AdS lengthscale at infinity) and fixed temperature $T$. The plot is quite similar to the 3-dimensional one \ref{Fig:CritAngleVsGamma}. Like in the 3-dimensional case, we focus on the negative $\gamma$ regime.

%
%
%
\section{Conclusions and future directions}\label{Sec:Conclusion}
In  this paper we obtained the  holographic entanglement entropy functional for a particular class of gravity  with tensor-scalar coupling, Horndeski gravity. We find that the  entanglement entropy receives a Wald-like contribution proportional to the gradient-square of the scalar field. We show that, as in Lewkowycz-Maldacena,  demanding that the divergence of the $zz$ component of the bulk equation of motion vanishes allows us to  identify the surface where to evaluate the entanglement functional. This surface turns out to be the one that minimizes said functional.
We also  pointed out  the existence of other divergences that deserve more study: quadratic divergences in other components of the equation of motion and subleading divergences in the $zz$ component.
As an application of the entanglement functional found, we present explicit minimal surfaces for black hole solutions and  show that they exhibit similar features to the ones observed in Schwarzchild-AdS: the connected surface ceases to exist for sufficiently large boundary region  and  there exist subdominant saddles. We also study the phase transition due to the exchange of dominance between  connected and disconnected surfaces. We show that the size of the entanglement plateau (at fixed temperature)  depends on the non-minimal coupling.

The thermal entropy of the   Horndeski  black holes we study here  was not well established. It has been previously investigated in the literature but  different methods of calculating it   yielded different results. We  used the entanglement entropy functional derived here to shed light on this issue. We identified an oversight in the literature  and determined  the correct thermal entropy. 

Let us conclude with some future directions.
\begin{itemize}
\item {\it Other divergences.} As previously mentioned in section \ref{Divergences}, the cancellation of the $T_{zz}$ divergence  implies that the entanglement functional should be evaluated on the surface that minimizes it. However, similar to what occurs in higher derivative theories, there are divergences in other components of the equations of motion. In \cite{Camps:2014voa} the authors showed that in the case of Gauss-Bonnet gravity these other divergences cancel if a more general ansatz is taken. This ansatz includes two types of new terms compared to (\ref{RNC}): terms that break replica-symmetry, and terms that can be gauged away at $\epsilon=0$ (but not at nonzero $\epsilon$). It is the latter kind of new terms that are responsible for the cancellation of subleading divergences in the case of Gauss-Bonnet gravity. It would be very interesting to investigate if  a similar cancellation occurs in the case of scalar-tensor gravities.

\item {\it Splitting problem.} As mentioned previously, the splitting problem has to do with the fact that the scalar field on the minimal surface splits into a sum of different contributions when we turn on $\epsilon$. Investigating this splitting pattern further is of interest. Such an investigation was carried out in the simpler case of dilaton gravity in \cite{Dong:2017xht} by solving the equation of motion at $\epsilon \approx 0$ near the tip of the cone. We also expect that a resolution of the splitting problem will shed some light on the problem of cancelling divergences of the equation of motion mentioned in the previous point.

\item {\it Field theory dual.} Identifying the precise dual of Horndeski gravity  is an open question that deserves study. In particular, carrying out the holographic renormalization programme for Horndeski gravity seems an important and attainable goal.
\item {\it Causal wedge.} The causal structure of Horndeski gravity has been extensively studied \cite{Minamitsuji:2015nca, Kobayashi:2014wsa}. In the  context of holography, it is understood that the RT surface should lie on the causal shadow in order for the boundary theory to be causally well defined. In \cite{Headrick:2014cta} it was proven that this is indeed the case if we consider Einstein gravity. It would be interesting to verify if this is also the case in Horndeski or more general scalar-tensor theories. This causality constraint could be used to rule out certain scalar-tensor theories from having QFT duals. 
\item{\it Conformally coupled theories} Conformally coupled black holes have been knnown for quite some time \cite{Martinez:1996gn}\cite{Henneaux:2002wm}\cite{Giribet:2014bva}\cite{Chernicoff:2016jsu}. It would be interesting to derive, following the same approach as in the present work, the entanglement functional relevant for those theories.
\item{\it Quantum corrections}  $1/N$ quantum corrections to the entanglement entropy involve  $S_{bulk}$ {\it i.e.} the entanglement entropy of the entanglement wedge with the rest of the spacetime, in a manner reminiscent of the generalized entropy of black holes. The scalar-tensor coupling will surely contribute to $S_{bulk}$.
\end{itemize}

\acknowledgments
We would like to thank Joan Camps for reading the manuscript and giving us useful comments. This material is based upon work supported by the National Science Foundation under Grant Number PHY-1620610 and was performed in part at Aspen Center for Physics, which is supported by National Science Foundation grant PHY-1607611. E.C. would like to thank Centro de Ciencias de Benasque Pedro Pascual for its hospitality while this work was completed. E. C. was also supported by Mexico's National Council of Science and Technology (CONACYT) grant CB-2014-01-238734 and by a grant from the Simons Foundation.

\appendix

\section{Minimal surfaces in 3d}\label{App:RTSurfaces}
We plot in Figure \ref{Fig:PlanarBHSurfaces} a few Ryu-Takayanagi surfaces for the 3-dimensional planar black hole for various values of $\gamma$ at fixed $g$, $\mu$ and $\beta$. In this appendix, we elaborate further on this plot. For the Einstein case $\gamma=0$, the surface (black curve in Figure \ref{Fig:PlanarBHSurfaces}) is given by the analytical expression (we set $g=\mu=1$):
\begin{equation}
z{(y)} = \sqrt{\frac{1-K\cosh{(2y)} }{1+K\cosh{(2y)} }}
\end{equation}
Here $K$ is an integration constant.\\
Although this may not be apparent from Figure \ref{Fig:PlanarBHSurfaces}, the near-boundary region of these surfaces is qualitatively different depending on whether $\gamma$ is negative or positive. For positive $\gamma$, the surface is always perpendicular to the boundary, while for negative $\gamma$ this is not the case: the surface does not approach the boundary at right angle for negative $\gamma$. This fact can appear surprising, but it comes from the contribution of the scalar field to the entanglement entropy: the $\gamma$ term is basically the norm-squared of the gradient of the scalar field on the surface, and this term occurs with a negative sign in the functional (\ref{Functional}).\\
Note that the norm-squared of any vector is positive in a Riemannian metric. Hence, the sign of the $\gamma$ term in the functional (\ref{Functional}) is the opposite of the sign of the coupling $\gamma$ itself. For negative $\gamma$, the functional is minimized if the magnitude of $h^{ij}\chi_{,i}\chi_{,j}$ is minimized on the surface. But since $\chi$ is only a function of $z$, the quantity $h^{ij}\chi_{,i}\chi_{,j}$ has maximal magnitude if the surface approaches the boundary perpendicularly. Therefore, in the case of negative $\gamma$, the surface should approach the boundary at some angle less than $\pi/2$ to keep the magnitude of $h^{ij}\chi_{,i}\chi_{,j}$ small. On the other hand, for positive $\gamma$, the functional becomes smaller if the magnitude of $h^{ij}\chi_{,i}\chi_{,j}$ becomes larger. In this case, the surface should approach the boundary perpendicularly because this is when $h^{ij}\chi_{,i}\chi_{,j}$ is maximal.\\
As for the global black hole, we present in the left panel of Figure \ref{Fig:GlobalBHSurfaces} the connected minimal surface for various boundary interval (from very small to the whole boundary circle) for a negative value of $\gamma$. Like the usual minimal surface of the Einstein-gravity BTZ black hole, the connected surface can wrap around the horizon all the way.

\begin{figure}
\centering
\begin{minipage}{.45\textwidth}
\centering
\includegraphics[width=0.6\linewidth]{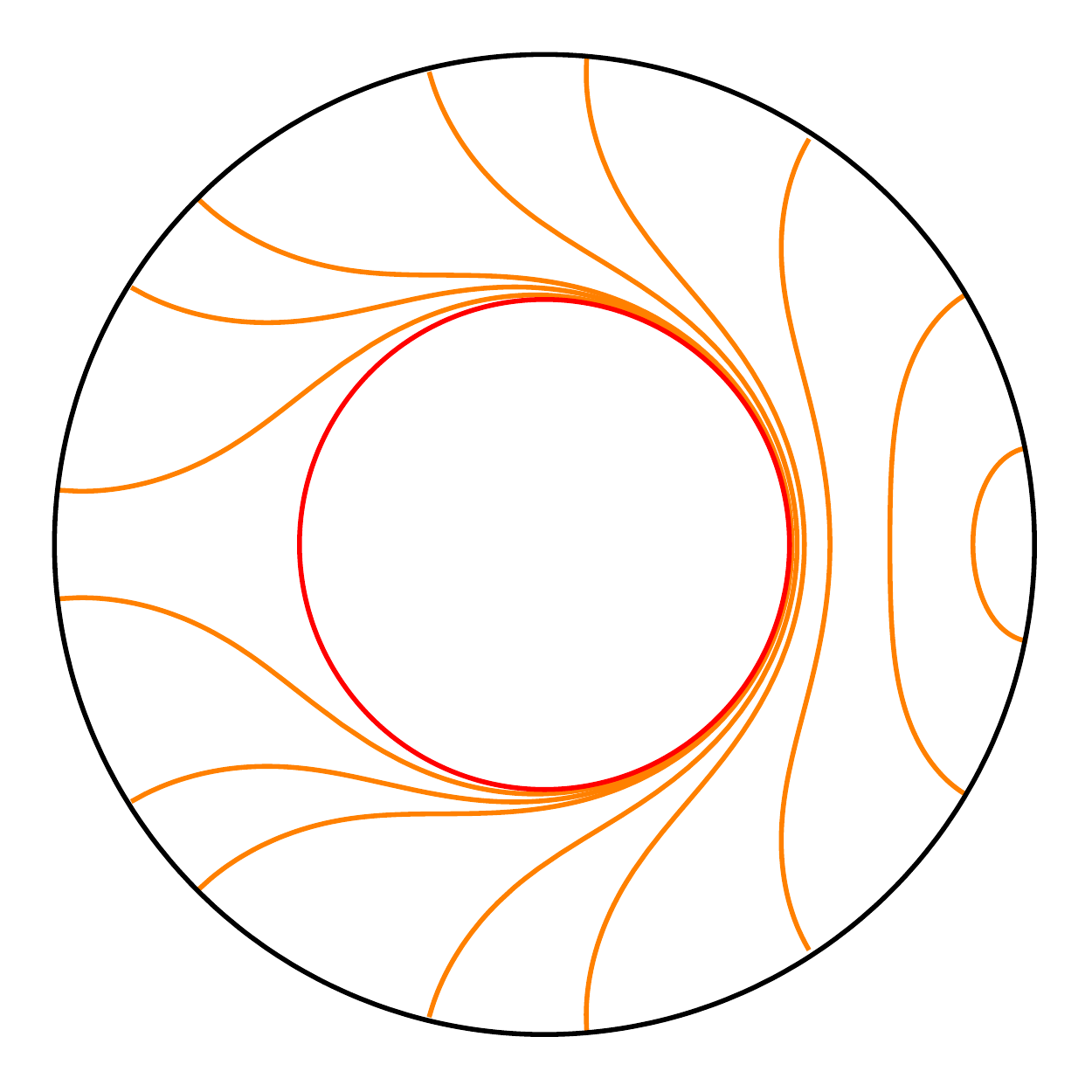}
\label{Fig:Global3dSurfaces}
\end{minipage}%
\begin{minipage}{.45\textwidth}
\centering
\includegraphics[width=0.6\linewidth]{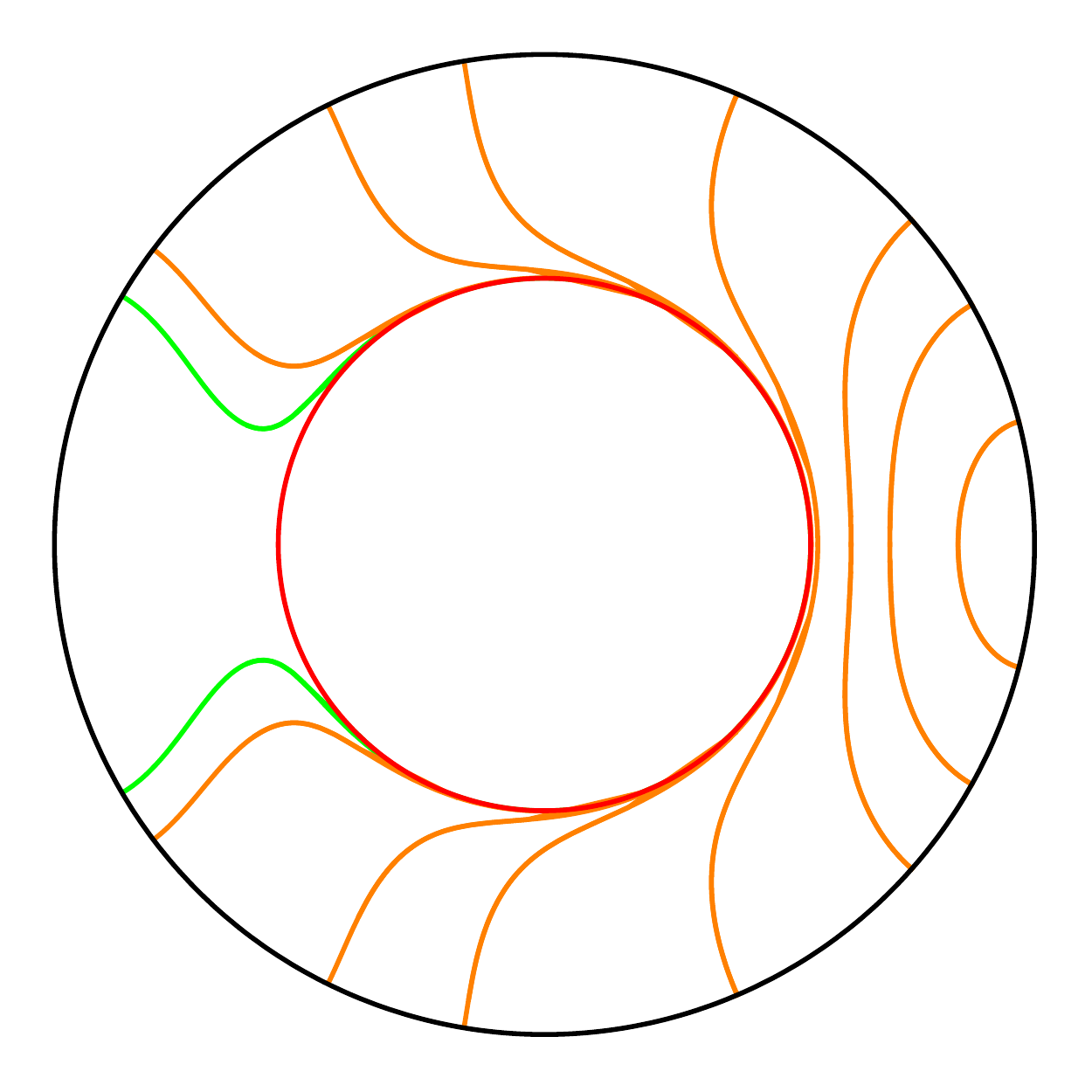}
\label{Fig:Global4dSurfaces}
\end{minipage}%
\end{figure}
\begin{figure}
\centering
\caption{Connected minimal surfaces. Left: $d=3$, $g=\mu=\beta=G=1$ and $\gamma=-0.1$. Right: $d=4$, $g=\beta=G=1$, $\gamma=-0.1$ and $\mu = 2.711055$. }
\label{Fig:GlobalBHSurfaces}
\end{figure}%

\section{Minimal surfaces in 4d}\label{App:RTSurfaces2}
\begin{figure}[h]
\centering
\includegraphics[width=0.3\textwidth]{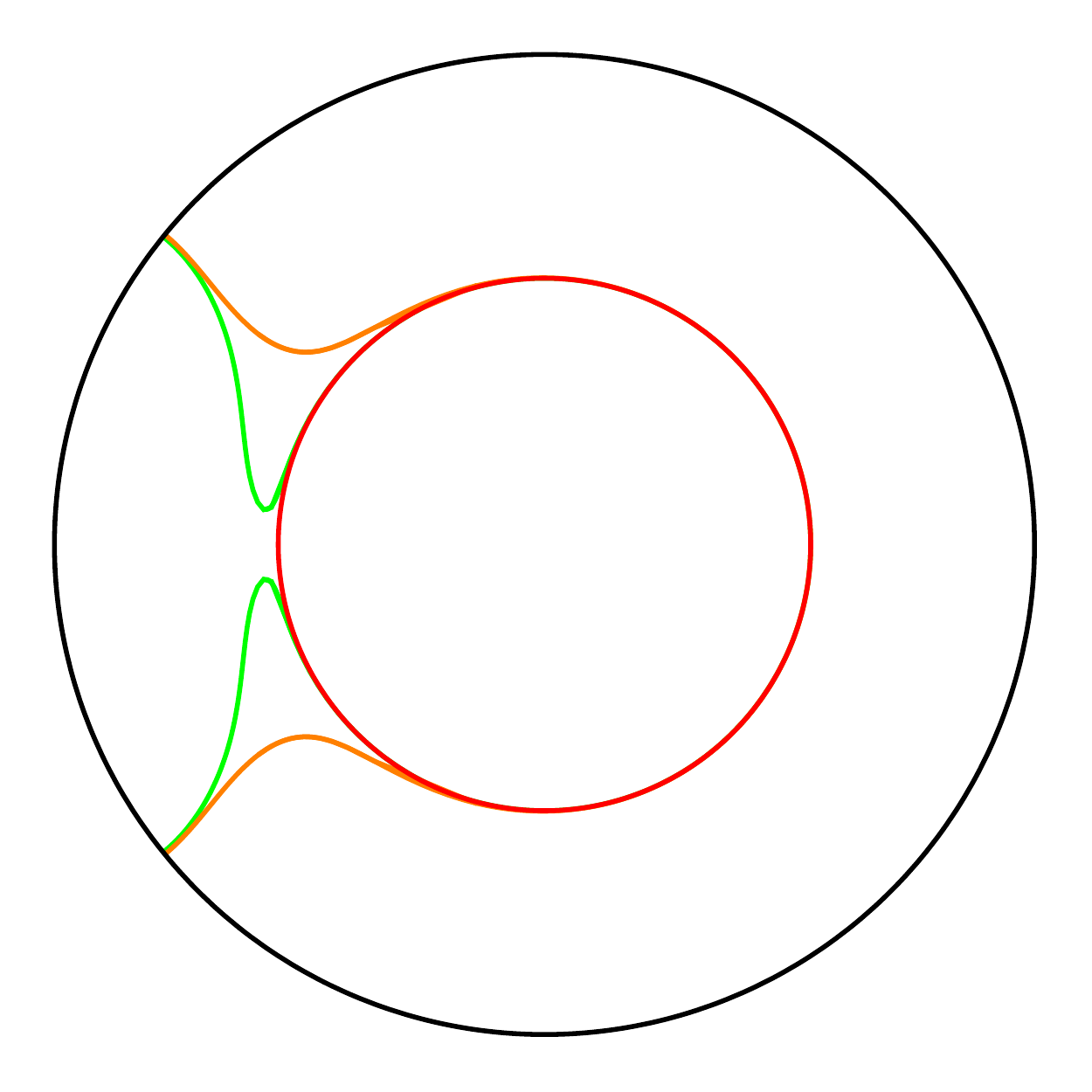}
\caption{A subdominant saddle (in green) and the minimal surface (in orange), $d=4$, $g=\beta=G=1$, $\gamma=-0.1$ and $\mu = 2.711055$.}
\label{Fig:Global4dSubdominant}
\end{figure}

In this appendix, we present the plots of the RT surfaces for the 3+1 dimensional (spherical) Horndeski black hole, with the boundary region taken to be a disk $\theta \leq \theta_{0}$. These surfaces behave qualitatively different from the ones in 2+1 dimensions in 2 ways:
\begin{itemize}
    \item The connected surface does not exist for all values of $\theta_{0}$. There exists a critical value $\theta_{m}$ such that no connected RT surface exists for $\theta_{0} > \theta_{m}$. In general, the threshold $\theta_{m}$ depends on the numerical values of the coupling constants, in particular $\gamma$. In the right-hand side of Figure \ref{Fig:GlobalBHSurfaces}, we plot a few connected RT surfaces for various sizes of the boundary region, up to the critical value $\theta_{m}$.

    \item There exists subdominant saddles which come closer to the horizon than the threshold surface $\theta_{0} = \theta_{m}$. For the same boundary region, we may have more than one RT surfaces: the dominant one and the subdominant one. We illustrate this in Figure \ref{Fig:Global4dSubdominant}.
\end{itemize}

We note that both features above (the existence of $\theta_{m}$ and of subdominant saddles) are also present in Einstein gravity, as explained in \cite{Hubeny:2013gta}.

\end{document}